\documentclass[prb,twocolumn,groupedaddress,showpacs]{revtex4}

\usepackage{graphicx}
\usepackage{dcolumn}
\usepackage{amssymb}
\usepackage{bm}
\usepackage{epsfig}

\begin{document}

\title{Transport properties of single channel quantum wires with an impurity:
\\Influence of finite length and temperature on average current and noise}

\author{Fabrizio~Dolcini$^1$, Bj\"orn~Trauzettel$^2$, In\`es~Safi$^2$, and
  Hermann Grabert$^1$}

\affiliation{${}^1$ Physikalisches Institut,
Albert-Ludwigs-Universit\"at,
79104 Freiburg, Germany\\
${}^2$Laboratoire de Physique des Solides, Universit\'e Paris-Sud,
91405 Orsay, France}

\date{Received 2 September 2004, published 13 April 2005}

\begin{abstract}
The inhomogeneous Tomonaga Luttinger liquid model describing an
interacting quantum wire adiabatically coupled to non-interacting
leads is analyzed in the presence of a weak impurity within the
wire. Due to strong electronic correlations in the wire, the
effects of impurity backscattering, finite bias, finite
temperature, and finite length lead to characteristic
non-monotonic parameter dependencies of the average current. We
discuss oscillations of the non-linear current voltage
characteristics that arise due to reflections of plasmon modes at
the impurity and quasi Andreev reflections at the contacts, and
show how these oscillations are washed out by decoherence at
finite temperature. Furthermore, the finite frequency current
noise is investigated in detail. We find that the effective charge
extracted in the shot noise regime in the weak backscattering
limit decisively depends on the noise frequency $\omega$ relative
to $v_F/gL$, where $v_F$ is the Fermi velocity, $g$ the Tomonaga
Luttinger interaction parameter, and $L$ the length of the wire.
The interplay of finite bias, finite temperature, and finite
length yields rich structure in the noise spectrum  which
crucially depends on the electron-electron interaction. In
particular, the excess noise, defined as the change of the noise
due to the applied voltage, can become negative and is
non-vanishing even for noise frequencies larger than the applied
voltage, which are signatures of correlation effects.
\end{abstract}

\pacs{71.10.Pm, 72.10.-d, 72.70.+m, 73.23.-b}

\maketitle
%%%%%%%%%%%%%%%%%%%%%%%%%%%%%%%%%%%%%%%%%%%%%%%%%
%%%%%%%%%%%%%%%%%%%%%%%%%%%%%%%%%%%%%%%%%%%%%%%%%
%%%%%%%%%%%%%%%%%%%%%%%%%%%%%%%%%%%%%%%%%%%%%%%%%
%%%%%%      I N T R O D U C T I O N       %%%%%%%
%%%%%%%%%%%%%%%%%%%%%%%%%%%%%%%%%%%%%%%%%%%%%%%%%
%%%%%%%%%%%%%%%%%%%%%%%%%%%%%%%%%%%%%%%%%%%%%%%%%
%%%%%%%%%%%%%%%%%%%%%%%%%%%%%%%%%%%%%%%%%%%%%%%%%
\section{Introduction}
Transport in Tomonaga-Luttinger liquid (TLL) systems has gained a
lot of attraction since the appearance of experimental
realizations of TLLs, such as cleaved edge overgrowth quantum
wires and single wall carbon nanotubes (SWNTs). However, a
quantitative comparison between theory and experiment often
suffers from the fact that the interaction parameter $g$ of the
TLL is not very well known and does not affect the DC conductance
and the shot noise of clean wires attached to Fermi liquid leads.
As a hallmark of the TLL model has served the power law dependence
of the nonlinear conductance of impure wires with respect to
temperature and/or applied voltage, which has been observed in
transport measurements in semiconductor quantum wires
\cite{tarucha,yacoby} as well as in nanotubes. \cite{bockrath}
However, some doubts in the determination of $g$ often remain,
since dynamical Coulomb blockade leads as well to a power law when
the system is embedded in an ohmic environment. \cite{devor90} It
has been shown recently that both the intrinsic interactions and
the environmental resistance enter on equal footing in $g$.
\cite{safi04} Thus, it is important to have alternative ways to
measure $g$ in order to validate the existence of TLL systems in
nature.

Crucial ingredients, which have been disregarded in earlier
theoretical work on transport in TLLs,
\cite{apel_rice,kane_fisher} are the finite wire length and the
role of the electron reservoirs. A first step towards a better
understanding of such systems has been taken by modelling the
reservoirs as one-dimensional non-interacting leads, introducing
the inhomogeneous TLL (ITLL) model.
\cite{ines_schulz,maslov_g,ponomarenko} It has been shown that the
DC conductance of the ITLL is independent of the electron
interaction strength. The analysis of this result has lead to a
new phenomenon: a momentum conserving reflection at the contacts
due to strong electronic correlations in the wire and their
absence in the leads. \cite{ines_schulz} This effect is similar to
the Andreev reflections at a metal-superconductor contact.
\cite{ines_ann} Apparently, a finite wire behaves as an
Andreev-type resonator for plasmon excitations: An electron
incident from a contact is transmitted in a series of current
spikes, which sum up to the flux of an electron, so that
ultimately the incident particle is perfectly transmitted. This
dynamics gives rise to nontrivial correlation functions.
\cite{ines_nato,furusaki_fil_fini,ines_ann,sandler} The ITLL model
has also been applied recently to study the effects of the Fermi
liquid leads on tunneling from an STM tip into a
nanotube.\cite{Martin} Apart from the ITLL model, the finite
length of an interacting quantum wire (QW) is also taken into
account in the approach of open boundary bosonization
\cite{fabri95}. However, there the wire is taken as a disconnected
object with sharp edges.

In the presence of an impurity in the QW, the ITLL model in its
standard form  \cite{ines_schulz,maslov_g,ponomarenko} can only be
treated perturbatively for weak (or strong) impurity strength. As
far as the calculation of the average current is concerned, this
problem can in principle be eluded by modelling the presence of
non-interacting leads through radiative boundary conditions.
\cite{egger} These have been combined with methods of
refermionization \cite{egg_gra} and integrable field theories
\cite{egger00} to obtain the exact current voltage
characteristics. The non-perturbative analysis of
Refs.~[\onlinecite{egg_gra,egger00}] is  restricted, up to now, to
an infinitely long wire, and it is currently not clear whether
this approach can also be applied to the calculation of current
fluctuations. For  a finite one-dimensional wire at low energies,
an alternative formulation of the radiative boundary conditions in
terms of operators instead of expectation values has been
proposed.\cite{ines_epjb} This approach  recovers the
quasi-Andreev reflections, and could perhaps also enable a
treatment of the noise.\cite{pham03}

Both the voltage dependence
of the differential conductance   and the frequency dependence of
the spectrum of the noise are expected to provide valuable
information on interaction effects in a finite length QW. In
Ref.~[\onlinecite{dolcini}] we have determined the nonlinear
current-voltage characteristics of a QW with an impurity for the
case of zero temperature, taking its finite length explicitly into
account through the ITLL model. The $I-V$ characteristics shows
interaction-dependent oscillations that are due to interference
effects between the Andreev-type reflections at the wire-lead
contacts and the backscattering at the impurity site. In the
present article, we analyze in detail how these oscillations are
modified at finite temperature. For a QW with applied voltage $V$
at temperature $T$ there is now apart from $eV$ and $k_B T$ a
third relevant energy scale $\hbar v_F/gL$, where the ballistic
frequency $v_F/gL$ is the ratio of the plasmon velocity $v_F/g$
and the length of the wire $L$. As an important consequence of
this new energy scale, the current as a function of $T$ at $V=0$
and the current as a function of $V$ at $T=0$ are not simply
related to each other by an interchange of $eV$ and $k_B T$ as it
is the case in the homogeneous TLL.

The main purpose of the present work, however, is to go beyond the
investigation of DC properties of ITLL systems by looking at the
finite frequency (FF) current noise. Much of the recent interest
in noise comes from the fact that shot noise may allow to
determine the effective charge that is backscattered off an
impurity in the QW. \cite{blanter00} Due to the dominance of $1/f$
noise at low frequencies, shot noise is never really measured at
zero frequency, but only down to the kHz range. This raises the
question on the influence of a finite measurement frequency on the
shot noise level. This issue has been  addressed theoretically for
a four-terminal fractional quantum Hall edge state geometry.
\cite{chamo96} However, the analysis was restricted to chiral TLLs
at zero temperature and infinite system size, which effectively
corresponds to a vanishing ballistic frequency $v_F/g L$. In
contrast, as will be shown in the present paper, for non-chiral
TLLs the ratio between $\omega$ and $v_F/gL$ crucially affects the
properties of FF noise. For this reason, it is essential to adopt
a model which  takes the finite length of the system  into
account. Here, we address this problem within the ITLL model.

From an experimental point of view, there is an advantage in
studying FF noise as compared to the AC conductance. While recent
theoretical work \cite{ines_schulz,ines_epjb,blanter} has
predicted that the AC conductance exhibits Andreev-type
oscillations, the high frequency range of these oscillations
cannot easily be explored in experiments. The problem is that the
frequency $\omega_{\rm AC}$ of the reservoir potential in AC
measurements must be low enough to ensure that inelastic processes
are sufficiently efficient to establish quasi-equilibrium
distributions as assumed in the theory. \cite{ines_epjb,blanter}
This means that $\omega_{\rm AC} \tau_{in}\ll 1$, where
$\tau_{in}$ is the characteristic time for inelastic
thermalization processes. However, the frequency $\omega$ that
appears in the expression of the FF noise derived below is not
limited by such assumptions. Furthermore, it is possible to
measure FF noise in a DC biased circuit.

In the regime $\omega \ll v_F/gL$, the shot noise of an ITLL
system in the presence of a weak backscattering potential has been
shown to be given by a $g$-independent classical Schottky formula
$S=2eI_{\rm BS}$, where $I_{\rm BS}$ is the backscattering
current. \cite{ponomarenko_sn,trauz_sn} Hence, the shot noise is
proportional to the electron charge $e$. In contrast, in the
homogeneous TLL model, the fractional charge extracted from the
ratio of the noise and the backscattering current in the weak
backscattering limit is given by $e^*=eg$ and thus  depends on the
interaction parameter $g$. \cite{kane_fisher_noise} This has
essentially been confirmed in shot noise measurements on
fractional quantum Hall edge state devices at filling fraction
$\nu=1/3$. \cite{depic,saminad} However, these systems are
described by the chiral TLL model, where right and left movers are
spatially separated and the effect of contact electrodes is quite
different from the non-chiral case discussed here.

Now, the question arises, if it is possible to observe the
fractional charge $e^*=eg$ also in QW realizations such as cleaved
edge overgrowth QWs or SWNTs. As we have shown
recently\cite{trauz04}, at frequencies $\omega \approx v_F/gL$,
the effective charge extracted by averaging the FF noise in ITLL
systems over a frequency range is indeed the quasiparticle charge
$e^*=eg$ and not the electron charge $e$. The origin of this
behavior is easily understood: On the one hand, for $\omega \ll
v_F/gL$, the noise probes current correlations on long time scales
allowing for a large number of scattering processes (at the
impurity and Andreev-type reflections at the boundaries to the
leads). Thus, the noise probes the current correlations of the
non-interacting leads with a complicated scatterer -- the
interacting QW and the impurity. On the other hand, in the case
$\omega \approx v_F/gL$, intrinsic spectral properties of the QW
determine the FF noise. Here, we go beyond the analysis of
Ref.~[\onlinecite{trauz04}] by explicitly considering the
influence of finite temperature on the equilibrium as well as
non-equilibrium current noise of a finite-length QW coupled to
electron reservoirs. Our general result for the FF noise of a wire
with arbitrary length $L$ describes the crossover between the shot
noise results for homogeneous TLL \cite{kane_fisher_noise} and
ITLL systems.\cite{ponomarenko_sn,trauz_sn} The observability of
the fractional charge backscattered at an impurity in a carbon
nanotube has also recently been discussed for a four-terminal
setup.\cite{Bena} However, the Hamiltonian used in Ref.\
[\onlinecite{Bena}] does not take the role of the Fermi liquid
leads fully into account and entails a factor of $g$ in the shot
noise strength even at $\omega \to 0$ in contrast to the findings
in Refs.\ [\onlinecite{ponomarenko_sn,trauz_sn}].

On the experimental side, there has been considerable progress in
the study of FF noise, recently. The FF noise of an electrically
driven two-state system has been measured, \cite{deblock} and the
high frequency current noise in a diffusive mesoscopic conductor
has been observed. \cite{schoelkopf} The low frequency regime has
also been explored experimentally: Shot noise has been measured in
bundles of carbon nanotubes \cite{roche02} and in SWNT.
\cite{kim03} However, the nature of the system in the former case
and the analysis of data in the latter case do not allow to
conclude whether the theoretical predictions
\cite{ponomarenko_sn,trauz_sn} on the shot noise in non-chiral TLL
systems have been confirmed or not, so that further work is
needed.

The article is organized as follows. In Sec.~\ref{sec_mod}, we
introduce the ITLL model. Then, in Sec.~\ref{sec_cur}, we first
discuss the influence of the finite length and electron-electron
interactions on the $I-V$ characteristics. Subsequently,  in
Sec.~\ref{sec_noi}, we present  results on the FF noise in ITLL
systems and discuss various limits, where we can make contact with
earlier work, in particular, the shot noise of the homogeneous TLL
model. Finally, we conclude in Sec.~\ref{sec_con}. Technical
details are given in the Appendices.

%%%%%%%%%%%%%%%%%%%%%%%%%%%%%%%%%%%%%%%%%%%%%%%%%
%%%%%%%%%%%%%%%%%%%%%%%%%%%%%%%%%%%%%%%%%%%%%%%%%
%%%%%%%%%%%%%%%%%%%%%%%%%%%%%%%%%%%%%%%%%%%%%%%%%
%%%%%%             M O D E L              %%%%%%%
%%%%%%%%%%%%%%%%%%%%%%%%%%%%%%%%%%%%%%%%%%%%%%%%%
%%%%%%%%%%%%%%%%%%%%%%%%%%%%%%%%%%%%%%%%%%%%%%%%%
%%%%%%%%%%%%%%%%%%%%%%%%%%%%%%%%%%%%%%%%%%%%%%%%%
\section{Model}
\label{sec_mod}

\begin{figure}
\vspace{0.3cm}
\begin{center}
\epsfig{file=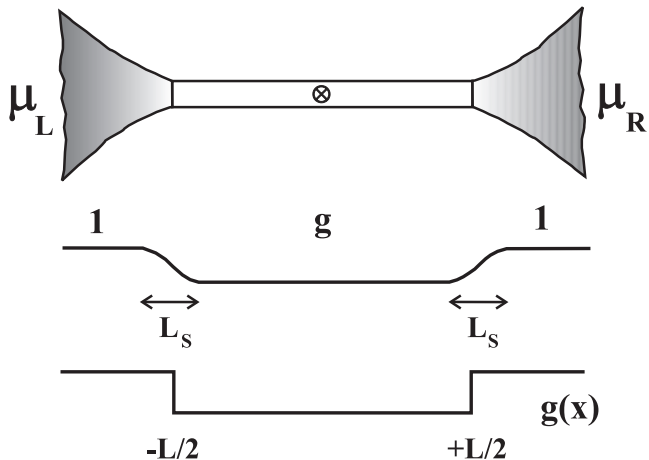,scale=0.45,=}
\caption{\label{setup} The upper part of the figure shows a QW
with an impurity adiabatically coupled to Fermi liquid leads. In
order to allow for a finite bias, the leads are held on different
electro-chemical potentials $\mu_L$ and $\mu_R$. The middle part
of the figure shows the actual variation of the TLL parameter $g$
along the wire-leads system in the ITLL model, and the lower part
of the figure its simplification under the assumption that
$\lambda_F \ll L_s \ll L$.}
\end{center}
\end{figure}

We model the physical system by the Hamiltonian
\begin{equation}
{\mathcal{H}} ={\mathcal{H}}_{0}  \, + \, {\mathcal{H}}_{B}  \, +
\, {\mathcal{H}}_{V} \; , \label{L}
\end{equation}
where ${\mathcal{H}}_{0}$ describes the interacting wire, the
leads and their mutual contacts, ${\mathcal{H}}_{B}$ accounts for
the electron-impurity interaction, and
 ${\mathcal{H}}_{V}$ contains the electrochemical bias applied to the wire.
Explicitly, we have
\begin{eqnarray}
{\mathcal{H}}_0 &=&\frac{\hbar v_F}{2}  \int_{-\infty}^{\infty}
 dx \left[ \Pi^2 + \frac{1}{g^2(x)}
(\partial _x\Phi )^2\right]  \, , \label{L0}  \\
{\mathcal{H}}_B &=& \lambda \cos{[\sqrt{4 \pi} \Phi(x_0,t)+2 k_F
x_0]} \label{LB} \; ,\\
{\mathcal{H}}_{V}  &=&    - \int_{-\infty}^{\infty}
\frac{dx}{\sqrt{\pi}} \, \mu(x) \,
\partial_x \Phi(x,t) \; . \label{LV}
\end{eqnarray}
Here, $\Phi(x,t)$ is the standard Bose field operator in
bosonization and $\Pi(x,t)$ its conjugate momentum density.
\cite{haldane_bosonisation} Eq.~(\ref{L0}) describes the
(spinless)  ITLL mentioned in the introduction, which is known to
capture the physical features of a QW with short-ranged (screened)
Coulomb interaction connected to metallic leads through adiabatic
contacts. The interaction parameter $g(x)$ is space-dependent and
its value is $g$ in the bulk of the wire, 1 in the bulk of the
leads, and supposed to change smoothly from 1 to $g$ at the
contacts. A schematic view of this model is shown in
Fig.~\ref{setup}. The variation of $g(x)$ is assumed to occur
within a characteristic smoothing length $L_s$. The profile of the
function $g(x)$ then identifies two energy scales $E_L=\hbar
v_F/gL$ and $E_s=\hbar v_F/L_s$, associated with the length $L$ of
the wire and the smoothing length~$L_s$. We assume here that $L_s
\gg \lambda_F$ where $\lambda_F$ is the electron Fermi wavelength.
Under these conditions, the electron-electron interaction in the
bosonized language remains quadratic in the field $\Phi$, and no
backscattering term arises at the wire-lead interfaces, as it is
expected for adiabatic contacts.

We also consider the situation $L_s \ll L$, so that $E_L \ll E_s$.
In this case the QW has a well defined length, as it is typically
the case for carbon nanotubes contacted  by metallic
leads.\cite{Nygard,Kong,liang01} The energy scale $E_s$ is then
much larger than the relevant energy scales that must be
considered to describe the effects of the finite wire length on
the average current and the noise. The precise form of the
variation of $g(x)$ at the contacts from $g$ to 1 is therefore not
physically relevant to our purposes, and it is consistent to
simplify the calculations by adopting a step-like function, namely
$g(x)=g$ if $x$ is in the wire (i.e. if $|x|<L/2$) and $g(x)=1$ if
$x$ is in the leads (i.e. if $|x|>L/2$), {\it cf.}
Fig.~\ref{setup}. We mention that the behavior at energies of
order $E_s$ has been addressed by Kleimann {\it et al.} in their
study of a one-dimensional quantum dot.\cite{kleimann}

The value $v_F$ appearing in Eq.~(\ref{L0}), although being of the
same order as the (equilibrium) bare Fermi velocity of the system,
may not exactly coincide with the latter. Indeed, as customary
when adopting a continuum limit of a lattice model, the actual
value of $v_F$ can be renormalized by the electron density as well
as by the interaction. \cite{haeusler} In a more detailed model of
the ITLL, one could also consider the plasmon velocity $v$ as an
independent parameter and replace $v_F/g$ by $v$ in
Eq.~(\ref{L0}). \cite{ines_schulz,ines_ann} Then, one would take
the possible renormalization of $v$ by irrelevant processes, such
as Umklapp scattering, into account. As our final results are
essentially unaffected  by such a sophistication of the model, we
will stick to the simple ITLL given by Eq.~(\ref{L0}) and assume
that the relation $v=v_F/g$ describes the interaction dependence
of the plasmon velocity.

Eq.~(\ref{LB}) is the $2k_F$ backscattering term at the impurity
site $x_0$, and introduces a strong non-linearity in the
field~$\Phi$. Near the low-energy fixed-point of the model this
term represents a relevant perturbation.  As mentioned in the
introduction, we will consider here the weak backscattering limit,
which amounts to treating Eq.~(\ref{LB}) as a perturbation. The
conditions on the coupling constant $\lambda$ in order for this
perturbative approach to be reliable have been discussed in
Ref.~[\onlinecite{dolcini}] and are recalled below. In passing, we
mention that impurity scattering also yields a forward scattering
term, which has been omitted, because it is unimportant for the
transport properties studied below.

%%%%%%%%%%%%%%%%%%%%%%%%%%%%%%%%%%%%%%%
%%%%%%%%%%%%%%%%%%%%%%%%%%%%%%%%%%%%%%%
%%%%%      FIGURE    1          %%%%%%
%%%%%%%%%%%%%%%%%%%%%%%%%%%%%%%%%%%%%%%
%%%%%%%%%%%%%%%%%%%%%%%%%%%%%%%%%%%%%%%
Finally, Eq.~(\ref{LV}) describes the coupling to the
electro-chemical bias due to the leads. In most experiments, leads
are normal 2D or 3D metals, and a detailed description of them
would require the standard Fermi liquid model. However, since we
are interested in properties of the {\it wire}, such a detailed
description of the leads would in fact be superfluous. One can
 account for their main effect, the applied bias voltage at the
contacts, by treating them as non-interacting systems ($g=1$), as
mentioned above. The only essential properties originating from
the Coulomb interaction that one needs to retain are i) the
possibility to shift the band-bottom of the leads, and ii)
electroneutrality in the leads. \cite{trauz_sn}

Therefore, the function $\mu(x)$ appearing in Eq.~(\ref{LV}),
which describes the externally tunable electro-chemical bias, is
taken as piecewise constant \cite{ines_ann}
\begin{equation}
\mu(x) = \left\{
\begin{array}{ll}
\mu_L & \mbox{for } x < -\frac{L}{2} \\
& \\
0 & \mbox{for }  |x| < \frac{L}{2}\\
& \\
\mu_R & \mbox{for }  x > + \frac{L}{2} \\
\end{array}
\right. \label{mu-profile}
\end{equation}
corresponding to an applied voltage
\begin{equation}
V \,\doteq \,(\mu_L-\mu_R) \, / \, e  \; . \label{def-voltage}
\end{equation}
We recall that in the absence of the non-linear term (\ref{LB}),
the current turns out to be exactly linear with respect to the
applied voltage $V$ (see e.g. Ref.~[\onlinecite{ines_schulz}]).

In bosonization, the current operator is related to the Bosonic
field $\Phi$ through
\begin{equation}
j (x,t) = \frac{e}{\sqrt{\pi}} \partial_t
\Phi(x,t) \; . \label{current} \\
\end{equation}
The finite frequency noise $S(x,y,\omega)$ is defined as
\begin{eqnarray}
S(x,y,\omega) = \int_{-\infty}^{\infty} dt e^{i\omega t}
\left\langle \left\{ \Delta j (x,t) , \Delta j(y,0) \right\}
\right\rangle \; , \label{noise}
\end{eqnarray}
where $\{ \, , \, \}$ denotes the anticommutator and $\Delta
j(x,t) = j(x,t) - \langle j(x,t) \rangle$ is the current
fluctuation operator.

%%%%%%%%%%%%%%%%%%%%%%%%%%%%%%%%%%%%%%%%%%%%%%%%%
%%%%%%%%%%%%%%%%%%%%%%%%%%%%%%%%%%%%%%%%%%%%%%%%%
%%%%%%%%%%%%%%%%%%%%%%%%%%%%%%%%%%%%%%%%%%%%%%%%%
%%%%%%       S E C T I O N     III        %%%%%%%
%%%%%%%%%%%%%%%%%%%%%%%%%%%%%%%%%%%%%%%%%%%%%%%%%
%%%%%%%%%%%%%%%%%%%%%%%%%%%%%%%%%%%%%%%%%%%%%%%%%
%%%%%%%%%%%%%%%%%%%%%%%%%%%%%%%%%%%%%%%%%%%%%%%%%

\section{Finite length effects in the current voltage characteristics}
\label{sec_cur}

In order to understand the behavior of  current fluctuations,
especially in the shot noise limit, it is crucial to first provide
a thorough description of the average current. As already
mentioned above, some results on the current at $T=0$ have been
published previously.\cite{dolcini} In the present section we
shall summarize these results, and, in addition, generalize them
to the case of finite temperature.

Since we explicitly take into account the finite  wire length $L$,
an additional energy scale comes into play, namely the ballistic
frequency
\begin{equation}
\omega_L = v_F/gL
\end{equation}
of the plasmonic excitations in the wire. The average value of the
current may be written as $I=I_0-I_{\rm BS}$, where $I_0$ is the
current the system would exhibit in the absence of the impurity,
and $I_{\rm BS}$ is the backscattering current. In
App.~\ref{app_keld} we show that for arbitrary impurity strength
and time-dependent applied voltage the backscattering current
takes the form
\begin{equation}
I_{\rm BS}(x,t) = -\frac{\hbar \sqrt{\pi}}{e^2}
\int_{-\infty}^{+\infty} \! \! \! \! dt' \sigma_0(x,t;x_0,t')
\left\langle j_B(x_0,t') \right\rangle_{\rightarrow} \, \,
\label{IBS_nonpert}
\end{equation}
Here $\sigma_0(x,t;y,t')$ is the local conductivity of the clean
wire (see App.~\ref{app_keld}, Eq.~(\ref{sigma0def}))
\cite{ines_schulz,ines_ann,ines_epjb,blanter}, and $j_B(x_0,t) $
is the ``backscattering current operator'', defined as
\begin{equation}
j_{B}(x_0,t) = - \frac{e}{\hbar} \frac{\delta
\mathcal{H}_B}{\delta \Phi(x_0,t)}[\Phi+A_0] \; \label{ib_def} .
\end{equation}
$A_0(x_0,t)$ is the field shift emerging when one gauges away the
applied voltage, as done in App.~\ref{app_keld}. As before, $x_0$
is the position of the impurity. Finally, $\langle \dots
\rangle_{\rightarrow}$ denotes an average with respect to the
Hamiltonian
${\mathcal{H}}_{\rightarrow}={\mathcal{H}}_{0}[\Phi]+{\mathcal{H}}_{B}[\Phi+A_0]$,
which includes the shift $A_0$ as defined in Eq.~(\ref{S_0B}). In
the case of a DC applied voltage $V$, $A_0(x_0,t)$ only depends on
time (see Eq.~(\ref{A_0_DC})). Then, $I_0=(e^2/h) V$ and $I_{\rm
BS}$ are independent of position and time.

Throughout this article we shall mostly deal with the limit of
weak backscattering at the impurity. It is well known that in this
regime  the shot noise is directly proportional to the
backscattering current $I_{\rm BS}$. It is thus worthwhile
focussing on the latter, rather than on the total current $I$. We
perform a perturbative expansion in the impurity strength
$\lambda$ appearing in Eq.~(\ref{LB}). This is well grounded
whenever either of the three energy scales $eV$, $k_B T$, or
$\hbar \omega_L$ is much larger than~$\lambda$.

It can be shown (see App.~\ref{app_keld}, Eq.~(\ref{Phi-mv-2}) for
details) that for a DC voltage Eq.~(\ref{IBS_nonpert}) reads to
leading order in $\lambda$
\begin{equation}
I_{\rm BS} = \frac{e \lambda^{2}}{4 \hbar^2}
\int_{-\infty}^{\infty} d t \, e^{i \omega_0 t} \sum_{s=\pm} s \,
e^{4 \pi {C}_0(x_0,st;x_0,0)} \, , \label{IBS}
\end{equation}
where $\omega_0=eV/\hbar$ is the characteristic frequency related
to the applied voltage $V$, and ${C}_0(x_0,t;x_0,0)$ is the
correlation function of the bosonic field $\Phi(x,t)$ in the clean
system (see Eq.~(\ref{Creg_def}) in App.~\ref{app_corr_fun}),
calculated at the impurity position~$x_0$. For arbitrary
interaction strength $0<g \le 1$, applied voltage $V$, and
temperature $T$, Eq.~(\ref{IBS}) can readily be evaluated
numerically.

\begin{figure}
\vspace{0.3cm}
\begin{center}
\epsfig{file=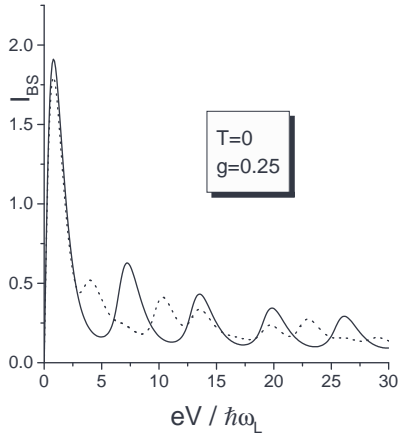,scale=0.3} \caption{\label{IBS_fig_T0}
Behavior of the backscattering current (in units of
$e(\lambda\omega_L^g/\omega_c^g)^2/\hbar^2 \omega_L$) as a
function of $V$, for two different impurity positions $\xi_0=0$
(solid line) and $\xi_0=\pm 0.2$ (dashed line).}
\end{center}
\end{figure}

Let us first recall the behavior at $T=0$. \cite{dolcini} In this
case, Eq.~(\ref{IBS}) for $I_{\rm BS}$  simplifies, because only
the term with `$s=+$'  contributes, and Eq.~(7) of
Ref.~[\onlinecite{dolcini}] is recovered. In Fig.~\ref{IBS_fig_T0}
we have plotted $I_{\rm BS}$ for the interaction strength $g=0.25$
(a typical value for SWNTs) and for two different impurity
positions, namely $\xi_0=0$ and $\xi_0=\pm 0.2$, where $\xi_0$ is
the {\it relative} position within the wire, i.e. $\xi_0=x_0/L$.
Notice that the current only depends on the absolute value
$|\xi_0|$ of the impurity shift off the center, and not on the
direction.
%%%%%%%%%%%%%%%%%%%%%%%%%%%%%%%%%%%%%%%
%%%%%%%%%%%%%%%%%%%%%%%%%%%%%%%%%%%%%%%
%%%%%        FIGURE   2          %%%%%%
%%%%%%%%%%%%%%%%%%%%%%%%%%%%%%%%%%%%%%%
%%%%%%%%%%%%%%%%%%%%%%%%%%%%%%%%%%%%%%%

The current exhibits oscillations as a function of the DC voltage
$V$. \cite{dolcini} The origin of these oscillations can be
understood from the following two effects: An adiabatic connection
between two TLLs with different interaction parameters (in our
case $g$ in the wire and $1$ in the leads), causes partial
reflections of plasmonic excitations at the contact region
(Andreev-type reflections); furthermore, in presence of an
impurity, an applied DC voltage can be regarded as a
time-dependent ``phase shifter'' located at the impurity site (see
the exponential factor $e^{i \omega_0 t}$ in Eq.~(\ref{IBS})).
This is due to the fact that the applied voltage can be gauged
away into a phase factor of the electron operators (see
App.~\ref{app_keld} Eqs.~(\ref{shift-in-field}), (\ref{ZB}), and
(\ref{Aeta}) for details).

In view of the above observations one can sketch the following
scenario: The plasmonic excitations are backscattered by the
impurity, driven towards the contacts where they exhibit further
partial Andreev-type reflections, and then come back to the
impurity with a phase difference $e^{i \omega_0 t_B}$, where $t_B$
is the ballistic time to propagate from the impurity to one of the
contacts and back. This phase difference is responsible for
interference effects, which can be tuned from constructive to
destructive nature by varying the DC voltage and/or the length of
the wire. This is basically the mechanism causing the oscillatory
behavior. The typical period of the oscillation of the current as
a function of the voltage is indeed $\Delta
V=2\pi\hbar\omega_L/e$. As can be seen from Fig.~\ref{IBS_fig_T0},
when the impurity is located away from the middle (i.e $|\xi_0|
\neq 0$), the ballistic times to reach the left and right contacts
are different, and therefore one observes two different
oscillation frequencies as in the dotted curve of
Fig.~\ref{IBS_fig_T0}.

Importantly, these current oscillations are a combined effect of
the impurity, the {\it finite-length}, and the {\it interaction}
in the wire, which causes the Andreev-type reflections at the
contacts. As soon as these reflections are suppressed (either
because the interaction in the wire is very weak, or the length is
sent to infinity) the oscillations disappear.

%%%%%%%%%%%%%%%%%%%%%%%%%%%%%%%%%%%%%%%
%%%%%%%%%%%%%%%%%%%%%%%%%%%%%%%%%%%%%%%
%%%%%       FIGURE    3          %%%%%%
%%%%%%%%%%%%%%%%%%%%%%%%%%%%%%%%%%%%%%%
%%%%%%%%%%%%%%%%%%%%%%%%%%%%%%%%%%%%%%%
\begin{figure}
\vspace{0.3cm}
\begin{center}
\epsfig{file=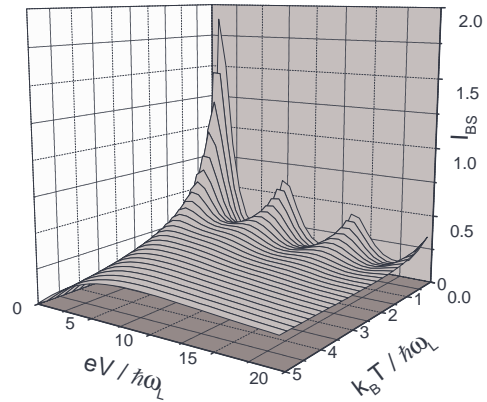,scale=0.35}
\caption{\label{IBS_fig_3D} The backscattering current (same units
as in Fig.~\ref{IBS_fig_T0}) as a function of voltage and
temperature.}
\end{center}
\end{figure}

Let us now consider the case of finite temperature $T>0$. Thermal
fluctuations are expected to induce decoherence in the
interference effects described above. In Fig.~\ref{IBS_fig_3D}, the
backscattering current $I_{\rm BS}$ is shown as a function of both
temperature and voltage, again for the case $g=0.25$. For
simplicity we have chosen an impurity located in the middle,
$\xi_0=0$. As one can see, when $k_BT$ becomes comparable with the
energy scale $\hbar\omega_L$ associated with the ballistic
frequency, the oscillations start to be smeared out. Importantly,
the existence of a third energy scale $\hbar \omega_L$ (apart
from $k_B T$ and $eV$) destroys the simple scaling behavior of
the current as a function of $V$ and/or $T$ known from the
homogeneous TLL. More precisely, the current as a function of $V$
at $T=0$ and the current as a function of $T$ at $V=0$ behave
very differently at intermediate energy scales, {\it cf.}
Fig.~\ref{IBS_fig_3D}. The same powerlaws for $I_{\rm BS}(V)$ at
$T=0$ as for $I_{\rm BS}(T)$ at $V=0$ only arise in the limit $L
\rightarrow \infty$, i.e. if $eV\gg \hbar \omega_L$ in the former
case and $k_B T\gg \hbar \omega_L$ in the latter case.

\begin{figure}
\vspace{0.3cm}
\begin{center}
\epsfig{file=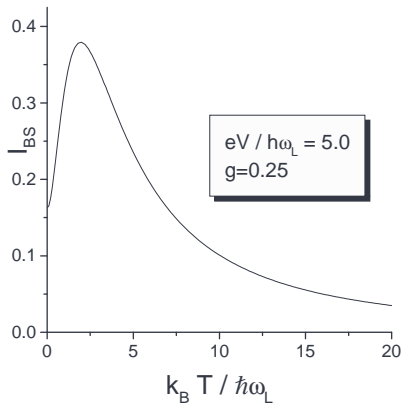,scale=0.3} \caption{\label{IBS_fig_V5}
The backscattering current (same units as in
Fig.~\ref{IBS_fig_T0}) as a function of temperature, for a voltage
value $eV=5 \, \hbar \omega_L$  corresponding to the first current
minimum at $T=0$ in Fig.~\ref{IBS_fig_T0}.  }
\end{center}
\end{figure}

Fig.~\ref{IBS_fig_V5} shows an interesting behavior of $I_{\rm
BS}$ at fixed voltage $V$ as a function of temperature that is due
to a combined effect of Andreev-type reflections and temperature
induced decoherence. In that figure, we focus on a voltage value
$V_{\rm min}$ corresponding to destructive interference at $T=0$,
i.e.\ to one of the current minima in Fig.~\ref{IBS_fig_T0} (solid
curve). If the temperature is increased, one first observes for
$k_B T \lesssim \hbar \omega_L$ an increase of $I_{\rm BS}$, due
to thermal decoherence of the destructive interference effect.
Then, for $k_B T \gg \hbar \omega_L$, $I_{\rm BS}$ decreases, so
that a maximum of $I_{\rm BS}$ as a function of $T$ is observed
when the voltage is near $V_{\rm min}$.

The above results have been obtained by means of a numerical
evaluation of Eq.~(\ref{IBS}). However, it is also possible to
derive analytical results for $I_{\rm BS}$, in particular in the
limit $eV \gg \hbar \omega_L$. In this case the two temperature
regimes $k_B T \ll \hbar \omega_L$ and $k_B T \gg
\hbar \omega_L$ can be distinguished.\\

(i) In the case $k_B T \ll \hbar \omega_L \ll eV$, it is
convenient to rewrite $I_{\rm BS}$ as
\begin{equation}
I_{\rm BS}\, = I^{\rm st}_{\rm BS}(V) \, [ 1\, + \, f_{\rm BS}(u,
\Theta,\xi_0)  ] , \label{IBS-factorized}
\end{equation}
where $I^{\rm st}_{\rm BS}$  is the leading order term and $f_{\rm
BS}$ gives the corrections. The former reads
\begin{equation}
I^{\rm st}_{\rm BS}(V)= \frac{e^2 V}{h}  \frac{\pi^2}{
\Gamma(2g)}\, \, \frac{ \left( \lambda/\hbar \omega_c
\right)^2}{\left( eV / \hbar \omega_c  \right)^{2(1-g)}}
\label{IBSinf}
\end{equation}
and is independent of the length $L$, of the temperature~$T$, and
of the impurity position $\xi_0$; it exhibits a power law behavior
as a function of the applied voltage $V$, in accordance with the
result for the homogeneous TLL. \cite{kane_fisher} In
Eq.~(\ref{IBSinf}), $\omega_c$ is the high-energy cutoff. As
usual, the cutoff dependence can be absorbed in an effective
impurity strength $\lambda^*=\hbar \omega_c (\lambda/\hbar
\omega_c)^{1/(1-g)}$, in terms of which Eq.~(\ref{IBSinf}) reads
\begin{equation}
I^{\rm st}_{\rm BS}(V)= \frac{e^2 V}{h}  \frac{\pi^2}{
\Gamma(2g)}\, \, \left( \frac{\lambda^{*}}{eV} \right)^{2(1-g)} .
\label{IBSinf-lambda-star}
\end{equation}
In contrast to the leading order term, the dimensionless
correction term $f_{\rm BS}$, which can be evaluated through an
asymptotic expansion of Eq.~(\ref{IBS}), explicitly depends on the
length $L$ of the wire, on the temperature $T$, and on the
position of the impurity $x_0$ through the dimensionless
parameters $u=eV/\hbar \omega_L$, $\Theta = k_B T /\hbar
\omega_L$, and $\xi_0=x_0/L$. We find that
\begin{equation} \label{DeltaIBS}
f_{\rm BS}(u, \Theta,\xi_0) \, = \, f^{\rm osc}_{\rm BS}(u,
\Theta,\xi_0) \, + \, \ldots ,
\end{equation}
where  $f^{\rm osc}_{\rm BS}$ describes the dominant oscillating
correction.  For $0 < |\xi_0| < 1/2$ the asymptotic expansion
yields
\begin{widetext}
\begin{equation}
f^{\rm osc}_{\rm BS}(u,\Theta,\xi_0)= \frac{2 \Gamma(2g)}{\Gamma(
g \gamma)} \sum_{s=\pm} {D}^{(1)}(s|\xi_0|;\Theta)\,\frac{\cos{[
(1-2s |\xi_0|) \, u \,- \, \pi g (1+s\gamma /2)]}}{u^{2g(1-
\gamma/2)}} \frac{(1+2s |\xi_0|)^{2 g \gamma(1-\gamma -
\gamma^2/2)} }{(1-2s |\xi_0|)^{g (2- \gamma)} (16 |\xi_0|)^{g
\gamma} } \label{asy-exp-xneq0}
\end{equation}
\end{widetext}
with
\begin{equation}
\gamma = \frac{1-g}{1+g} \label{gamma}
\end{equation}
and ${D}^{(1)}(\pm|\xi|;\Theta)$ is a numerical factor  of order
unity, explicitly given in Eq.~(\ref{D^1}). We mention here that a
similar asymptotic expansion has been worked out in
Ref.~[\onlinecite{ponomarenko_tunnel}], where the crossover from
TLL to Fermi liquid behavior has been studied for a tunnel
barrier.

Notice that Eq.~(\ref{asy-exp-xneq0}) is singular for $\xi_0
\rightarrow 0$. This is actually only a mathematical problem: When
the distances of the impurity from the two contacts become equal,
pairs of poles in the  correlation function merge. The full
expression (\ref{IBS}) for the current is however perfectly
regular, and one can indeed still calculate $f^{\rm osc}_{\rm BS}$
for $\xi_0=0$ obtaining
\begin{equation} \label{asy-exp-xeq0}
f^{\rm osc}_{\rm BS}(u,\Theta,0) = \frac{2 \Gamma(2g)}{\Gamma( g
\gamma)} {D}^{(2)}(\Theta)\,\frac{\cos{ [u \,- \, \pi g
(1+\gamma)]}}{u^{2g(1- \gamma)}} ,
\end{equation}
where ${D}^{(2)}(\Theta)$ is a numerical coefficient of order
unity,
whose detailed expression is given in Eq.~(\ref{D^2}).\\

\noindent (ii) In the regime $\hbar \omega_L \ll  \{k_B T,eV\}$,
$I_{\rm BS}$ can be well approximated by
\begin{eqnarray} \label{IBSasym-reg2}
I_{\rm BS} &\simeq & \frac{e}{h} \frac{(2 \pi)^{2 g}}{2 \Gamma(2
g)} k_B T \left(\frac{\lambda^{*}}{k_B T} \right)^{2(1-g)}
 \nonumber \\
& \times& \sinh(\frac{eV}{2 k_B T}) \, |\Gamma(g+i \frac{eV}{2 \pi
k_B T})|^2 .
\end{eqnarray}
Hence, if $\hbar \omega_L$ is the smallest of all relevant energy
scales, the oscillatory dependence of $I_{\rm BS}$ on $V$
disappears, and the current is independent of the length of the
wire, and, in particular, of the impurity position. Indeed, in
this regime the thermal decoherence time is much shorter than the
wire ballistic time; thus, the backscattering by the impurity
(occurring in the bulk of the wire) cannot be influenced by the
physics at the contacts. The result (\ref{IBSasym-reg2}) coincides
with the prediction for a homogeneous system (see e.g.
Ref.~[\onlinecite{kane_fisher_3term}]), provided that the voltage
is  rescaled $V \rightarrow gV$. Indeed, it is well known that  in
the homogeneous TLL an effective voltage $gV$ appears\cite{NOTA},
yielding, for instance, an interaction dependent conductance in
the case of a clean wire, which is not correct for the system with
leads under investigation.\cite{ines_schulz,maslov_g,ponomarenko}
Finally, we mention that when the interaction is switched off  ($
g \rightarrow 1$), the temperature-dependent contribution to
$I_{\rm BS}$ in Eq.~(\ref{IBSasym-reg2}) vanishes by virtue of the
identity (\ref{Gamma1+iX}). This is expected for non-interacting
electrons where $I_{\rm BS}$ is a linear function
of the applied voltage.\\
%%%%%%%%%%%%%%%%%%%%%%%%%%%%%%%%%%%%%%%
%%%%%%%%%%%%%%%%%%%%%%%%%%%%%%%%%%%%%%%
%%%%%      FIGURE    4          %%%%%%
%%%%%%%%%%%%%%%%%%%%%%%%%%%%%%%%%%%%%%%
%%%%%%%%%%%%%%%%%%%%%%%%%%%%%%%%%%%%%%%
%%%%%%%%%%%%%%%%%%%%%%%%%%%%%%%%%%%%%%%
%%%%%%%%%%%%%%%%%%%%%%%%%%%%%%%%%%%%%%%
%%%%%         FIGURE   5         %%%%%%
%%%%%%%%%%%%%%%%%%%%%%%%%%%%%%%%%%%%%%%
%%%%%%%%%%%%%%%%%%%%%%%%%%%%%%%%%%%%%%%
\begin{figure}
\begin{center}
\epsfig{file=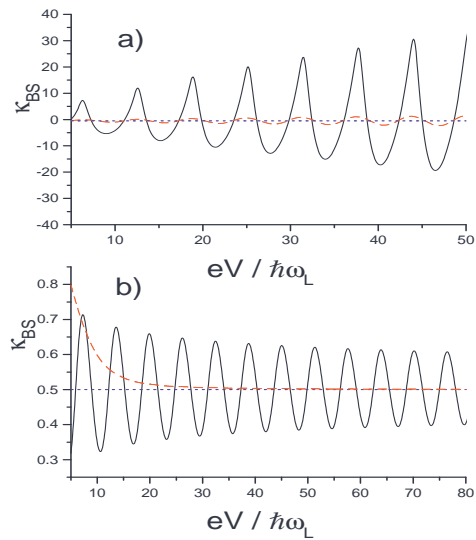,height=8cm,width=7cm}
\caption{\label{figM1_1} The dimensionless quantity $\kappa_{\rm
BS}(V)$ defined in Eq.~(\ref{kappa_def}) as a function of the
voltage, for a centered impurity $\xi_0=0$: a) shows the case of
strong interaction $g=0.25$, while b) depicts the case of weak
interaction $g=0.75$. The solid lines refer to zero temperature
and the dashed lines to the dimensionless temperature
$\Theta=k_BT/\hbar\omega_L=2$. The function $\kappa_{\rm BS}(V)$
oscillates around the value $2g-1$, indicated by the dotted line,
which corresponds to the power law exponent of $I_{\rm BS}$ in the
homogeneous TLL model. }
\end{center}
\end{figure}
%%%%%%%%%%%%%%%%%%%%%%%%%%%%%%%%%%%%%%%
%%%%%%%%%%%%%%%%%%%%%%%%%%%%%%%%%%%%%%%
%%%%%         FIGURE   6         %%%%%%
%%%%%%%%%%%%%%%%%%%%%%%%%%%%%%%%%%%%%%%
%%%%%%%%%%%%%%%%%%%%%%%%%%%%%%%%%%%%%%%
\begin{figure}
\begin{center}
\epsfig{file=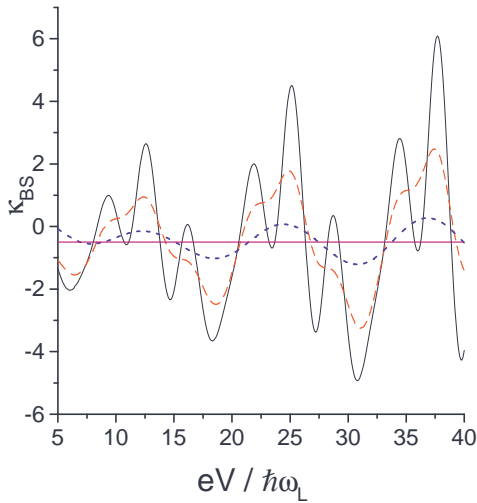,height=7cm,width=7cm}
\caption{\label{figM1_b} $\kappa_{\rm BS}(V)$  as a function of
the voltage for an off-centered impurity $\xi_0=0.25$, and for
strong interaction strength $g=0.25$. The three different curves
refer to three different values of the dimensionless temperature
$\Theta=k_B T/\hbar \omega_L$: $\Theta=0.5$ (solid), $\Theta=1$
(dashed) and $\Theta=3$ (dotted). The two frequencies mentioned in
the text are in this case $\omega_{L_1}/\omega_L=2/3$ and
$\omega_{L_2}/\omega_L=2$. For low temperatures the voltage
dependence is affected by Andreev-type reflections at both
contacts. For higher temperatures interference effects only
survive for  Andreev-type reflections at the closer contact, so
that only the frequency $\omega_{L_2}$ is present. The average
value of the oscillations is $2g-1$ (horizontal solid line),
whereas the period of the oscillations $\Delta V= 2
\pi\hbar\omega_L /e(1-2 |\xi_0|)$ reveals the position of the
impurity. }
\end{center}
\end{figure}

We conclude this section by discussing the behavior of a
dimensionless quantity that characterizes the current voltage
characteristics
\begin{equation}
\kappa_{\rm BS}(V) = \frac{V}{ I_{\rm BS}} \frac{d I_{\rm BS}}{d
V}  \, . \label{kappa_def}
\end{equation}
If the wire is assumed to be infinitely long, $I_{\rm BS}$ is
simply given by the right-hand side of Eq.~(\ref{IBSasym-reg2});
in this case, and in particular in the low temperatures regime
$k_B T \ll eV$, $\kappa_{\rm BS}$  is actually independent of the
voltage and equals $2g-1$, the power law exponent of $I_{\rm BS}$
in the homogeneous TLL model.
\\In contrast, for a finite length wire, $\kappa_{\rm BS}$ acquires a non trivial
voltage dependence, even at low temperatures. In particular, if
$k_B T \lesssim \hbar \omega_L$, $\kappa_{\rm BS}(V)$ exhibits an
oscillating behavior around $2g-1$, due to the Andreev-type
reflections discussed previously. An average of $\kappa_{\rm
BS}(V)$ over a sufficiently large range of voltages thus allows
for an estimate of $g$ from the current-voltage characteristics.
Higher temperatures $k_B T \gg \hbar \omega_L$ affect the
coherence of the Andreev-type processes, and the amplitude of the
oscillations is suppressed. On the other hand, the amplitude of
the oscillations also depends on the interaction strength. For the
case of a centered impurity, $\kappa_{\rm BS}$ is depicted in
Fig.~\ref{figM1_1}, both for strong and weak interaction strength.
Interestingly, with increasing voltage the amplitude of the
oscillations of $\kappa_{\rm BS}(V)$  increases for strong
interaction, and decreases for weak interaction. There is a
'critical' value $g=(1+\sqrt{17})/8 \simeq 0.64$ at which this
change of behavior occurs. This can be determined analytically
from the asymptotic expansion
(\ref{DeltaIBS})-(\ref{asy-exp-xeq0}) by requiring that the
exponent $2g(1-\gamma)$ of the power law in the denominator of
Eq.~(\ref{asy-exp-xeq0}) equals 1.

Fig.~\ref{figM1_b} refers to  the case of  an off-centered
impurity; in this case the behavior of $\kappa_{\rm BS}(V)$ is
characterized by two underlying frequencies $\omega_{L_1}$ and
$\omega_{L_2}$ ($\omega_{L_1} < \omega_{L_2}$), related to the
times needed by a plasmon excitation to travel from the impurity
to the contacts (see Eq.~(\ref{asy-exp-xneq0})). Thus, at low
temperatures $k_B T \ll \{ \hbar \omega_{L_1},\hbar \omega_{L_2}
\}$ the function $\kappa_{\rm BS}(V)$ exhibits two periods. Higher
temperatures introduce again decoherence which suppresses the
oscillations. However, in the temperature window $\hbar
\omega_{L_1} \lesssim k_B T \lesssim \hbar \omega_{L_2}$, only
coherence effects related to  the Andreev-type reflections from
the farther contact will be destroyed, whereas the ones related to
the closer contact are only weakly affected. Therefore
$\kappa_{\rm BS}(V)$ will exhibit oscillations with one period
only. In this regime, a graph of $\kappa_{\rm BS}(V)$ allows to
determine: i) the value of $g$ from the  average $2g-1$, and ii)
the position of the impurity from the period $\Delta V = \pi \hbar
v_F / eg (L/2-|x_0|)$ of the oscillations as a function of the
voltage.
%%%%%%%%%%%%%%%%%%%%%%%%%%%%%%%%%%%%%%%%%%%%%%%%%
%%%%%%%%%%%%%%%%%%%%%%%%%%%%%%%%%%%%%%%%%%%%%%%%%
%%%%%%%%%%%%%%%%%%%%%%%%%%%%%%%%%%%%%%%%%%%%%%%%%
%%%%%%       S E C T I O N     IV         %%%%%%%
%%%%%%%%%%%%%%%%%%%%%%%%%%%%%%%%%%%%%%%%%%%%%%%%%
%%%%%%%%%%%%%%%%%%%%%%%%%%%%%%%%%%%%%%%%%%%%%%%%%
%%%%%%%%%%%%%%%%%%%%%%%%%%%%%%%%%%%%%%%%%%%%%%%%%

\section{Finite frequency current noise}
\label{sec_noi}

We now turn to the FF current noise, defined previously in
Eq.~(\ref{noise}). The  noise at finite frequency $\omega$, finite
temperature $T$, finite bias $V$, and finite length $L$ will be
analyzed  in the presence of a weak backscatterer in the QW.
Mostly, this is done by numerical integration of a general
expression derived below. However, analytical results are
discussed for the case of thermal equilibrium and for the far from
equilibrium shot noise limit, when the applied voltage is much
larger than all other relevant energy scales.

A detailed derivation of the full expression for the FF noise in
the presence of an impurity is given in App.~\ref{app_noise}.
Here, we just mention the final result. In the presence of an
impurity, there is a contribution to the noise due to the
partitioning of the current at the backscatterer; as a
consequence, the noise can be written as
\begin{equation}
S(x,y,\omega)= S_0(x,y,\omega) \, + \, S_{\rm imp}(x,y,\omega)\, ,
\label{S0+Simp}
\end{equation}
where the first part, $S_0(x,y,\omega)$, is the current noise in
the absence of a backscatterer which will be   thoroughly
discussed in the following subsection. In contrast, $S_{\rm imp}$
is the supplementary  noise due to the impurity, which naturally
splits into two parts,
\begin{equation} \label{ff_result}
S_{\rm imp} (x,y,\omega)=S_A(x,y,\omega)+S_C(x,y,\omega) \; ,
\end{equation}
namely a contribution $S_A(x,y,\omega)$ related to the Fourier
transform of the anticommutator of the backscattering current
operator $j_B$, and a contribution $S_C(x,y,\omega)$ related to
the time-retarded commutator of $j_B$. Notice that, in order to
make the notation for $S$ lighter, we have suppressed the
temperature $T$ and voltage $V$ arguments, on which the noise
depends in general.

The first term in Eq.~(\ref{ff_result}) can be written in a
suitable way as
\begin{eqnarray} \label{S_A}
S_A(x,y,\omega) &=& \frac{1}{4\pi} \left(\frac{h}{e^2}\right)^2 \times \\
&& \sigma_0(x,x_0,\omega) f_A(x_0,\omega) \sigma_0(x_0,y,-\omega)
\nonumber
\end{eqnarray}
with
\begin{equation}
f_A(x_0,\omega) = \int_{-\infty}^\infty dt \, e^{i \omega t}
\left\langle \left\{ \Delta j_B(x_0,t), \Delta j_B(x_0,0) \right\}
\right\rangle_{\rightarrow} , \label{s_A}
\end{equation}
where $\Delta j_B(x_0,t) = j_B(x_0,t) - \langle j_B(x_0,t)
\rangle_\rightarrow$. Here, $j_B(x_0,t)$ is the backscattering
current operator defined in Eq.~(\ref{ib_def}), while
$\sigma_0(x,y;\omega)$ is the local conductivity of the clean
system discussed in the following subsection. $S_A(x,y,\omega)$ is
the dominant contribution to the noise out of equilibrium, i.e.~if
$eV \gg \{\hbar \omega_L, \hbar \omega, k_BT \}$. At zero
temperature and frequency, $S_A(x,y,\omega)$ is even the only
non-vanishing part of the noise and commonly called {\em shot
noise}. The shot noise is independent of the position, as will be
demonstrated below.

The second part, $S_C(x,y,\omega)$, is given by Eq.~(\ref{SCapp}).
Using the results in Apps. A and B, it can also be expressed as
\begin{widetext}
\begin{eqnarray} \label{S_C}
S_C(x,y,\omega) = \frac{h}{2 e^4 \omega} \Bigl\{ S_0(x,x_0,\omega)
f_C(x_0,-\omega) \sigma_0(x_0,y,-\omega) - S_0(y,x_0,-\omega)
f_C(x_0,\omega) \sigma_0(x_0,x,\omega) \Bigr\}
\end{eqnarray}
\end{widetext}
with
\begin{equation}
f_C(x_0,\omega) = \int_0^\infty dt \left( e^{i \omega t}-1 \right)
\left\langle \left[ j_B(x_0,t),j_B(x_0,0) \right]
\right\rangle_{\rightarrow} \; . \label{s_C}
\end{equation}
In the following subsections we discuss the equilibrium noise, the
non-equilibrium noise, and their difference, the excess noise.

%%%%%%%%%%%%%%%%%%%%%%%%%%%%%%%%%%%%%%%%%%%%%%%%%
%%%%%%%%%%%%%%%%%%%%%%%%%%%%%%%%%%%%%%%%%%%%%%%%%
%%%%%%%%%%%%%%%%%%%%%%%%%%%%%%%%%%%%%%%%%%%%%%%%%
\subsection{Equilibrium noise}

\subsubsection{Equilibrium noise in the clean system}

In the absence of a backscatterer, the noise is just given by the
first term $S_0$ of Eq.~(\ref{S0+Simp}). $S_0(x,y,\omega)$ is
directly connected to the Fourier transform of the anticommutator
$\tilde{{\mathcal{C}}}_0^K(x,y,\omega)$  of the bosonic phase
field $\Phi(x,t)$ (see Eq.~(\ref{S0app})). The properties of the
correlation function (\ref{corr-step1}) at finite temperature
allow to relate $S_0$ also to the conductivity through the
relation \cite{ponom96}
\begin{equation}
S_0(x,y,\omega)= 2 \hbar \omega \coth \left( \frac{\hbar
\omega}{2k_BT} \right) \Re [\sigma_0(x,y,\omega)] \, ,
\label{s0thermal}
\end{equation}
where $\Re$ denotes the real part. The latter equation is known as
the fluctuation-dissipation theorem (FDT). For its derivation, we
used that the conductivity can be expressed through the retarded
correlation function by the Kubo formula
\begin{equation} \label{sig0omega}
\sigma_0(x,y,\omega) =  \frac{2 e^2}{h} \omega
\tilde{C}_0^R(x,y,\omega) \; ,
\end{equation}
where $\tilde{C}_0^R(x,y,\omega)$ is given by Eq.~(\ref{cret}) in
combination with Eq.~(\ref{cfourier}). Eq.~(\ref{s0thermal}) holds
for $x=y$ in general, and also for $x \neq y$ if the unperturbed
Hamiltonian preserves time-reversal symmetry, as  in our case. The
relation (\ref{s0thermal}) holds for any temperature $T$, noise
frequency $\omega$, and system size $L$. This means that the
information that can be gained from the equilibrium noise is fully
contained in the AC conductivity. Notice that, by its definition
(\ref{sig0omega}), the AC conductivity $\sigma_0(x,y,\omega)$ is
voltage independent. Moreover, it can easily be shown to be also
temperature independent. This is a direct consequence of the fact
that the model Hamiltonian ${\mathcal{H}}_0$, Eq.~(\ref{L0}),
which describes an interacting QW without impurity, is quadratic
in the boson fields (see App.~\ref{app_corr_fun} for details).
Thus, in the absence of a backscatterer, the only temperature
dependence of the noise comes from the factor $\coth (\hbar
\omega/2k_BT )$ in Eq.~(\ref{s0thermal}). In order to determine
$S_0$, we are left with the analysis of the $x$, $\omega$, and $L$
dependence of the real part of the AC conductivity.

At very high frequencies, the FF noise becomes sensitive to its
point of measurement \cite{trauz_ff}, therefore, it is important
to discuss for which range of measurement positions the results of
our model will be reliable. As described in Sec.~\ref{sec_mod}, we
model the electron reservoirs by non-interacting one-dimensional
QWs. This is legitimate as long as we are not more than an
inelastic scattering length $l^{R}_{in}$ away from the contacts.
$l^{R}_{in}$ is the typical length scale at which inter-channel
scattering occurs in the reservoirs. Importantly, $l^{R}_{in}$ is
not necessarily of the same magnitude as the inelastic scattering
length in the wire $l^{W}_{in}$, which, of course, has to be
larger than $L$, in order to be in the ballistic regime. In some
situations, for instance, in cleaved edge overgrowth quantum
wires, we expect $l^{R}_{in} \approx l^{W}_{in}$, whereas in other
situations  we expect $l^{R}_{in} \ll l^{W}_{in}$, for instance,
in carbon nanotubes contacted by gold electrodes.

On the other hand, the model adopts a step-like profile at the
 contacts for the interaction strength $g(x)$  (see
Fig.~\ref{setup}). Therefore, if $d_x$ is the distance between the
measurement point (in the lead) and the closer contact, $d_x$
should not be smaller than the smoothing length $L_s$. Hence, we
can state that our model is reliable if the noise is measured at a
point $x$ in the range $L_s \lesssim d_x \lesssim l^{R}_{in}$.

For simplicity, we will now concentrate on the local FF noise,
i.e. we will put $x=y$ in Eq.~(\ref{s0thermal}). For a point $x$
in one of the leads it can be shown \cite{ines_ann} that
\begin{equation} \label{sigma0outside}
\sigma_0\left(x,x,\omega \right)= \frac{e^2}{h} \left( 1 + \gamma
\frac{2i \sin(\omega/\omega_L) e^{i2 \omega \delta_x /g \omega_L}
}{e^{-i \omega/\omega_L}-\gamma^2 e^{i\omega/\omega_L}} \right) \;
,
\end{equation}
where
\begin{equation}
\delta_x=(|x|-\frac{L}{2})/L = |\xi|-\frac{1}{2} \label{delta_x}
\end{equation}
is the distance $d_x$, in units of the wire length $L$. As before,
we use the notation $\xi = x/L$ and $\omega_L = v_F/gL$. The real
and imaginary parts of Eq.~(\ref{sigma0outside})
 read
\begin{widetext}
\begin{eqnarray}
\Re [\sigma_0\left(x,x,\omega \right)] &=& \frac{e^2}{h} \left[ 1
- \sin \left( 2  \omega \delta_x/g \omega_L \right) h_1( \omega/
\omega_L) - \cos \left(2  \omega \delta_x /g \omega_L \right)
h_2(\omega/\omega_L) \right] ,\label{realsigma0} \\
\Im [\sigma_0\left(x,x,\omega \right)] &=& \frac{e^2}{h} \left[
\cos \left( 2  \omega \delta_x/g \omega_L \right) h_1( \omega/
\omega_L) - \sin \left(2  \omega \delta_x / g \omega_L \right)
h_2(\omega/\omega_L) \right]  \label{imsigma0}
\end{eqnarray}
\end{widetext}
with the functions
\begin{eqnarray}
h_1(w) &=& \gamma(1-\gamma^2) \frac{\sin(2 w)}{1-2 \gamma^2 \cos(2
w)+\gamma^4} , \\
h_2(w) &=& \gamma(1+\gamma^2) \frac{1-\cos(2 w)}{1-2 \gamma^2
\cos(2 w)+\gamma^4} .
\end{eqnarray}
%%%%%%%%%%%%%%%%%%%%%%%%%%%%%%%%%%%%%%%
%%%%%%%%%%%%%%%%%%%%%%%%%%%%%%%%%%%%%%%
%%%%%        FIGURE    7         %%%%%%
%%%%%%%%%%%%%%%%%%%%%%%%%%%%%%%%%%%%%%%
%%%%%%%%%%%%%%%%%%%%%%%%%%%%%%%%%%%%%%%
\begin{figure}
\begin{center}
\epsfig{file=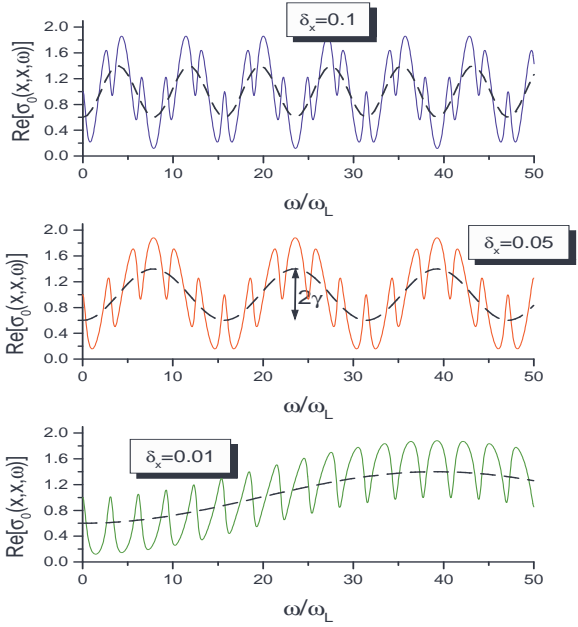,height=9cm,width=8cm}
\caption{\label{cond-fig1} The real part of the conductivity
(\ref{realsigma0}) of a clean wire (in units of $e^2/h$), is
plotted as a function of $\omega$, for three different positions
$x$ of the measurement point, and for interaction strength
$g=0.25$; the parameter $\delta_x=(|x|-L/2)/L$ defines the
distance of $x$ to the closer contact. One can see the oscillatory
behavior characterized by the two periods $\Delta \omega_1$ and
$\Delta \omega_2$, with $\Delta \omega_1 \ll \Delta \omega_2$. As
the point $x$ gets closer to the contact the beating period
$\Delta \omega_2$ tends to infinity. The dashed curve results from
an averaging over the rapidly oscillating component with period
$\Delta \omega_1$ according to Eq.~(\ref{realsigma0-aver}). The
oscillation amplitude of the dashed curve is $\gamma$, i.e.\ it
allows to determine the interaction strength $g$ from
Eq.~(\ref{gamma}). }
\end{center}
\end{figure}
The spatial dependence of the noise follows from
Eqs.~(\ref{s0thermal}) and (\ref{realsigma0}). In the regime
$\omega \ll \omega_L$, the spatial dependence vanishes. In
particular, if $k_B T \gg \hbar |\omega|$, we obtain the thermal
or {\em Johnson-Nyquist noise}
\begin{equation} \label{john-nyqu-noise}
S_0(x,x,0) = 4 k_B T \frac{e^2}{h} \; ,
\end{equation}
whereas if $k_B T \ll \hbar |\omega|$ we recover the quantum noise
or the {\em zero point fluctuations}
\begin{equation} \label{ZPF}
S_0(x,x,\omega) = 2 \hbar |\omega| \frac{e^2}{h} \; .
\end{equation}
Hence, if $\omega \ll \omega_L$, the noise does not contain any
information about the electron-electron interaction in the finite
length QW.

More interesting features appear when $\omega$ is comparable or
much bigger than the ballistic frequency $\omega_L$. In this
regime Eq.~(\ref{realsigma0}) is oscillating as a function of
$\omega$ with two characteristic periods $\Delta \omega_1$ and
$\Delta \omega_2$: the former, appearing through $h_{1,2}$, is
related to the ballistic frequency $\Delta \omega_1= \pi
\omega_L$; the latter depends on the measurement point and reads
$\Delta \omega_2= \pi g \omega_L/\delta_x$. Typically one has
$\Delta \omega_1 \ll \Delta \omega_2$, since $\delta_x \ll 1 $;
one therefore expects a sort of beating behavior, as shown in
Fig.~\ref{cond-fig1} for three different values of~$x$.

The dashed curves represent the function
\begin{equation}
\Re [\sigma_0^{\rm slow}\left(x,x,\omega \right)] \, = \,
\frac{e^2}{h} \left[ 1-\gamma \cos(2 \omega \delta_x/g
\omega_L)\right] \, ,\label{realsigma0-aver}
\end{equation}
 obtained by averaging out
the fast oscillations, i.e.\ by replacing the functions
$h_{1,2}(\omega/\omega_L)$ in (\ref{realsigma0}) by their average
values $\langle h_{1} \rangle_{\Delta \omega_1} = 0$ and $\langle
h_{2} \rangle_{\Delta \omega_1} = \gamma$, where the averaging is
defined as
\begin{equation}
\langle f(\omega/\omega_L) \rangle_{\Delta
\omega_1}=\frac{1}{\Delta \omega_1} \int_0^{\Delta \omega_1} f
(\omega/\omega_L) d \omega . \label{average-fast}
\end{equation}
Interestingly, for any $\delta_x \neq 0$, $\Re [\sigma_0^{\rm
slow}\left(x,x,\omega \right)]$ in units of $e^2/h$ oscillates
around 1 with amplitude $\gamma$ which is directly connected to
the interaction strength $g$ (see Eq.~(\ref{gamma})).

For the special case $\delta_x=0$ (i.e.\ if one could ideally
measure the noise {\it at the contacts}), Eq.~(\ref{realsigma0})
becomes strictly periodic in $\Delta \omega_1$ (the period $\Delta
\omega_2 \rightarrow \infty$), and the averaging procedure
(\ref{average-fast}) directly yields
\begin{equation}
\Re \left[\sigma_0^{\rm slow}\left(\frac{L}{2},\frac{L}{2},\omega
\right)\right]
 \, = \, \frac{e^2}{h} (1-\gamma)
\, = \, \frac{e^2}{h} g_c \; , \label{realsigma0-aver-cont}
\end{equation}
where $g_c=2 g/(1+g)$  is the effective TLL interaction parameter
at a point contact between a Fermi liquid (with interaction
parameter $g=1$) and a TLL (with interaction
parameter $g$). \cite{ines_schulz,sandler}\\

\begin{figure}
\begin{center}
\epsfig{file=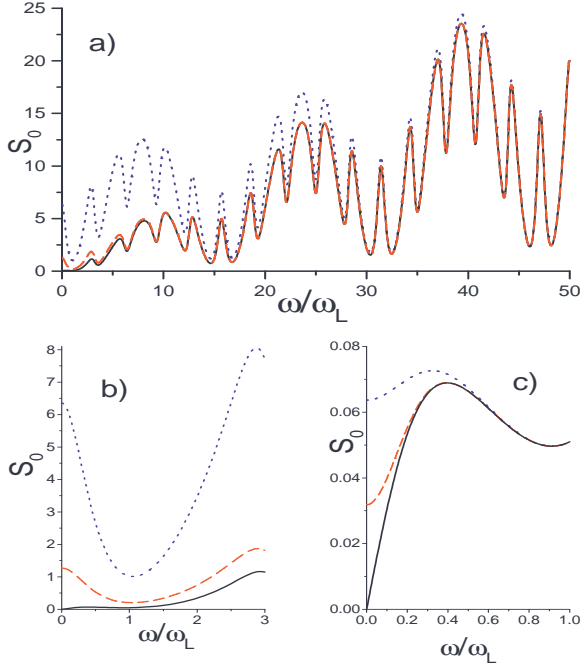,height=10cm,width=9cm}
\caption{\label{noise0} The noise in the absence of the impurity
(in units of $e^{2}\omega_L$), as a function of $\omega/\omega_L$,
for the measurement position $\delta_x=0.05$. a) the whole range
of frequencies for three values of the dimensionless temperature
$\Theta=k_B T/\hbar \omega_L$ (solid curve $\Theta=0$, dashed
curve $\Theta=2$, and dotted curve $\Theta=10$). b) enlargement
for  the range of low frequencies $\omega \in [0,3 \omega_L]$. c)
the case of low temperatures (solid curve $\Theta=0$, dashed curve
$\Theta=0.05$, and dotted curve $\Theta=0.1$).}
\end{center}
\end{figure}

The frequency dependence of the noise $S_0(x,x,\omega)$, related
to $\Re [\sigma_0]$ through Eq.~(\ref{s0thermal}), is shown in
Fig.~\ref{noise0} for the measurement position $\delta_x=0.05$
(see Eq.~(\ref{delta_x})), and for three different values  of
temperature $\Theta=k_B T/\hbar \omega_L$ ($\Theta=0,2,10$). The
noise is shown only for positive $\omega$ since $S_0$ is symmetric
in $\omega$ by definition. In Fig.~\ref{noise0}a one can see that
at high frequencies $\omega \gg \omega_L$ the noise exhibits a
linear growth modulated by oscillations with period
$\Delta\omega_1=\pi \omega_L$. Fig.~\ref{noise0}b is a blow-up of
Fig.~\ref{noise0}a at small frequencies, showing for  $\omega \ll
\omega_L$ the Johnson-Nyquist noise (\ref{john-nyqu-noise}).
Finally, Fig.~\ref{noise0}c refers to low temperatures
($\Theta=0,0.05,0.1$) displaying   at $T=0$ and small frequencies
the  zero-point noise (\ref{ZPF}).
%%%%%%%%%%%%%%%%%%%%%%%%%%%%%%%%%%%%%%%
%%%%%%%%%%%%%%%%%%%%%%%%%%%%%%%%%%%%%%%
%%%%%       FIGURE    8          %%%%%%
%%%%%%%%%%%%%%%%%%%%%%%%%%%%%%%%%%%%%%%
%%%%%%%%%%%%%%%%%%%%%%%%%%%%%%%%%%%%%%%

In the introduction, we have mentioned that in terms of comparison
between theory and experiments, it is more suitable to determine
$\Re [\sigma_0 \left(x,x,\omega \right)]$ from a DC noise
measurement (through the FDT relation (\ref{s0thermal})) than from
a direct AC conductivity measurement.  Here, we have shown that by
averaging $\Re [\sigma_0\left(x,x,\omega \right)]$ over a
frequency range $[0, \pi \omega_L]$ local features of the wire
emerge, and the TLL parameter $g$ becomes accessible. This
statement will be further supported by the results presented
 in Sec.~\ref{sec_non_eq}.
%%%%%%%%%%%%%%%%%%%%%%%%%%%%%%%%%%%%%%%%%%%%%%%%%
%%%%%%%%%%%%%%%%%%%%%%%%%%%%%%%%%%%%%%%%%%%%%%%%%
%%%%%%%%%%%%%%%%%%%%%%%%%%%%%%%%%%%%%%%%%%%%%%%%%

\subsubsection{Equilibrium noise in presence of an impurity}
\label{sec-IV-A-2} In the presence of an impurity the FDT relating
noise~$S$ and local conductivity~$\sigma$ still holds. Explicitly
one has
\begin{equation}
\left. S(x,y,\omega) \right|_{V=0}= 2 \hbar \omega \coth \left(
\frac{\hbar \omega}{2k_BT} \right) \left. \Re [\sigma(x,y,\omega)]
\right|_{V=0} , \label{FDT}
\end{equation}
where the general expressions for the conductivity and the noise
are provided in App.~\ref{app_noise} (see Eq.~(\ref{sigmadef}) and
Eqs.~(\ref{SABC})--(\ref{SCapp}), respectively). From
Eq.~(\ref{FDT}) we observe that the noise is real, even for $x
\neq y$; this is due to the time-reversal symmetry of the
Hamiltonian and   the equilibrium condition.

The identity (\ref{FDT}) can be verified using
Eq.~(\ref{s0thermal}) and recalling that {\it at equilibrium} the
following properties hold: (i) the average values $\langle
\ldots\rangle_{\rightarrow}$ are evaluated with respect to the
Hamiltonian (\ref{S_0B}); (ii) $\langle j_B(x_0,t)
\rangle_\rightarrow = 0$; (iii) the functions $f_A$ and $f_C$
defined in Eqs. (\ref{s_A}) and (\ref{s_C}) are related by the
equation
\[
\left. f_A(x_0,\omega) \right|_{V=0} = 2 \coth\left( \frac{\hbar
\omega}{2 k_B T} \right) \Re \left[\left. f_C(x_0,\omega)
\right|_{V=0} \right] \; .
\]
Before providing explicit results in the weak backscattering
limit, we wish to emphasize two general aspects. First, in
contrast to the clean case, in the presence of an impurity the
conductivity depends on temperature; as a consequence, the
temperature-dependence of the noise is no longer simply given by
the $\coth$ factor in Eq.~(\ref{FDT}), and it will be analyzed
below.

Secondly, in the static limit $\omega \rightarrow 0$, one recovers
the  relation
\begin{equation} \label{highTresult}
\left. S(x,y,0) \right|_{V=0} = 4 k_B T G_\lambda   \; ,
\label{static=cond}
\end{equation}
where $G_\lambda$ is the conductance in the presence of the
impurity, defined as
\begin{equation}
G_\lambda \doteq  \left. \frac{d \langle j(\mathbf{x}) \rangle}{d
V} \right|_{V = 0} \; , \label{Gdef}
\end{equation}
which is independent of $\mathbf{x}=(x,t)$ but depends on the
temperature $T$. A direct calculation, using Eq.~(\ref{sigmadef})
in the limit $\omega \rightarrow 0$, shows that $G_\lambda=
\frac{e^2}{h}(1-{\mathcal R}_\lambda)$, where
\begin{equation}
{\mathcal R}_\lambda =   \frac{i}{2 e^2} \int_{0}^{\infty}  dt  \,
t \, \left. \langle [j_B(x_0,t), j_B(x_0,0)] \rangle \right|_{V=0}
\,  \;   \label{Rlambda}
\end{equation}
is an effective reflection coefficient. The behavior of ${\mathcal
R}_\lambda$   has been analyzed previously.
\cite{ines_ann,ines_nato} It has been shown that, in the regime
$k_B T \gg \hbar \omega_L$, ${\mathcal R}_\lambda$ shows a typical
TLL powerlaw behavior with respect to the temperature~$T$,
recovering the result for the homogeneous TLL. \cite{kane_fisher}
In contrast, in the regime $k_B T \ll \hbar \omega_L$, ${\mathcal
R}_\lambda $ shows a powerlaw behavior with respect to the length
$L$ of the QW.

In order to prove the relation (\ref{static=cond}) between the
static limit $\omega \rightarrow 0$ of the noise $S$ and the
conductance $G_\lambda$, we have calculated the equilibrium noise
from Eqs.~(\ref{S_A}), (\ref{S_C}) and (\ref{s0thermal}), and
compared the resulting expression with Eq.~(\ref{Rlambda}). \\
%%%%%%%%%%%%%%%%%%%%%%%%%%%%%%%%%%%%%%%
%%%%%%%%%%%%%%%%%%%%%%%%%%%%%%%%%%%%%%%
%%%%%        FIGURE   9          %%%%%%
%%%%%%%%%%%%%%%%%%%%%%%%%%%%%%%%%%%%%%%
%%%%%%%%%%%%%%%%%%%%%%%%%%%%%%%%%%%%%%%
\begin{figure}
\begin{center}
\epsfig{file=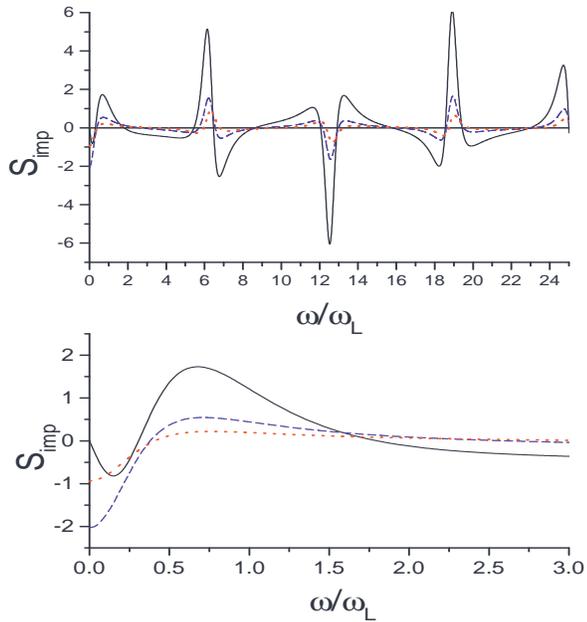,height=9.5cm,width=8.5cm}
\caption{\label{noise_eq}  The impurity noise $S_{\rm imp}$ at
equilibrium (in units of $e^2 \omega_L
(\lambda^{*}/\hbar\omega_L)^{2(1-g)}$), as a function of
$\omega/\omega_L$, for the measurement position $\delta_x=0.05$,
the impurity in the center ($\xi_0=0$), and three different values
of the dimensionless temperature $\Theta=k_B T / \hbar \omega_L$
(solid curve $\Theta=0$, dashed curve $\Theta=2$, and dotted curve
$\Theta=10$). Lower graph: Enlargement for low frequencies. Finite
temperature influences significantly the slope of the noise near
$\omega = 0$.}
\end{center}
\end{figure}

Let us now consider the case of finite frequency. The equilibrium
noise can be obtained from the conductivity through
Eq.~(\ref{FDT}). In particular, in the weak backscattering limit
an expansion of the conductivity in powers of $\lambda$ gives to
leading order
\begin{equation}
\sigma(x,y,\omega)=\sigma_0(x,y,\omega)+ \sigma_{\rm
BS}(x,y,\omega) \label{sig12}
\end{equation}
where
\begin{eqnarray}\nonumber
\sigma_{\rm BS}(x,y,\omega)= - \frac{2}{\hbar \omega} \left(
\frac{\pi \lambda}{e}\right)^2 \sigma_0(x,x_0,\omega)
\sigma_0(x_0,y,\omega) \\
\label{sigbs} \times \int_0^\infty dt \, (e^{i \omega t}-1) \left(
\sum_{s=\pm} s \, e^{4 \pi C_0(x_0,s t;x_0,0)} \right) .
\end{eqnarray}
We shall analyze here the local noise $S(x,x,\omega)$ for a
point~$x$ located in the leads. Comparing Eqs.~(\ref{FDT}) and
(\ref{S0+Simp}), one can easily recognize that the first term on
the r.h.s. of Eq.~(\ref{sig12}) yields the noise $S_0$ of the
clean wire, whereas the second term determines the supplementary
noise $S_{\rm imp}$ due to the backscattering at the impurity.
Since the former has been described in the previous subsection, we
shall focus now on the latter. The frequency spectrum of $S_{\rm
imp}$ is shown in Fig.~\ref{noise_eq} for three different values
of the dimensionless temperature $\Theta=k_BT/\hbar \omega_L$. The
upper graph of Fig.~\ref{noise_eq} shows that $S_{\rm imp}$
oscillates around $S_{\rm imp}=0$ with  pronounced spikes of
either sign, which are washed out when the temperature is
increased. While $S_{\rm imp}$ can be negative, the full noise
(\ref{noise}) is  non-negative. The lower graph of
Fig.~\ref{noise_eq} zooms into the region of low frequencies and
illustrates that, at finite temperature, the value of $S_{\rm
imp}$ at $\omega =0$ is negative. This is in accordance with
Eq.~(\ref{highTresult}), since the presence of the impurity
reduces the  conductance $G_{\lambda}$ with respect to the clean
case.
%%%%%%%%%%%%%%%%%%%%%%%%%%%%%%%%%%%%%%%
%%%%%%%%%%%%%%%%%%%%%%%%%%%%%%%%%%%%%%%
%%%%%        FIGURE  10          %%%%%%
%%%%%%%%%%%%%%%%%%%%%%%%%%%%%%%%%%%%%%%
%%%%%%%%%%%%%%%%%%%%%%%%%%%%%%%%%%%%%%%
\begin{figure}
\begin{center}
\epsfig{file=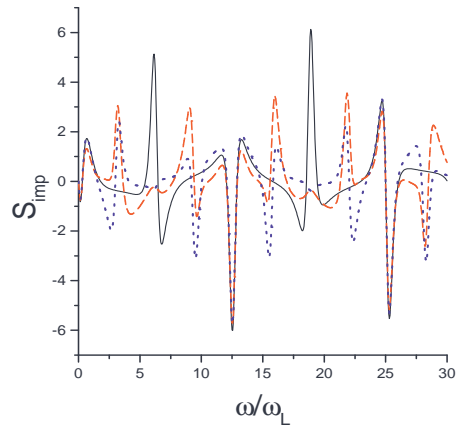,height=6cm,width=7cm}
\caption{\label{noise_eq_2} The zero temperature impurity noise
$S_{\rm imp}$ at equilibrium (in units of $e^2 \omega_L
(\lambda^{*}/\hbar\omega_L)^{2(1-g)}$) as a function of
$\omega/\omega_L$, for the measurement position $\delta_x=0.05$,
and interaction parameter $g=0.25$. Shown are  three cases of the
impurity position: $\xi_0=0$ (solid curve) a centered impurity,
$\xi_0=0.25$ (dashed curve) an impurity off-centered by 1/4
towards the direction of the measurement point, and $\xi_0=-0.25$
(dotted curve)  an impurity off-centered by 1/4 opposite to the
direction of the measurement point.}
\end{center}
\end{figure}
%%%%%%%%%%%%%%%%%%%%%%%%%%%%%%%%%%%%%%%
%%%%%%%%%%%%%%%%%%%%%%%%%%%%%%%%%%%%%%%
%%%%%        FIGURE   11         %%%%%%
%%%%%%%%%%%%%%%%%%%%%%%%%%%%%%%%%%%%%%%
%%%%%%%%%%%%%%%%%%%%%%%%%%%%%%%%%%%%%%%
\begin{figure}
\begin{center}
\epsfig{file=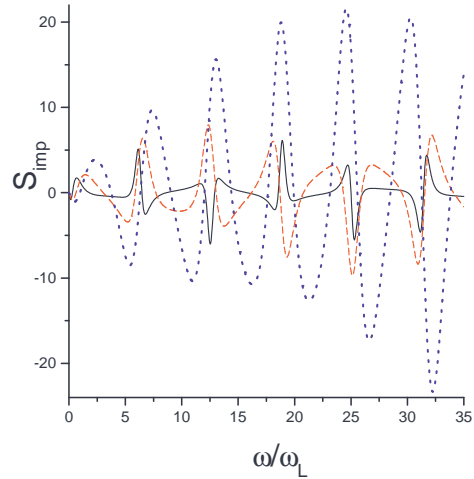,height=7cm,width=7cm}
\caption{\label{noise_eq_3} The zero temperature impurity noise
$S_{\rm imp}$ at equilibrium (in units of $e^2 \omega_L
(\lambda^{*}/\hbar\omega_L)^{2(1-g)}$) as a function of
$\omega/\omega_L$, for the measurement position $\delta_x=0.05$,
in the case of a centered impurity ($\xi_0=0$) and for three
different values of the interaction strength: $g=0.25$ (solid
curve), $g=0.50$ (dashed curve), and $g=0.75$ (dotted curve).}
\end{center}
\end{figure}

Fig.~\ref{noise_eq_2}   shows how the spikes of $S_{\rm imp}$
depend on the position of the impurity. Differently from the
average current shown in Fig.~\ref{IBS_fig_T0}, the current
fluctuations depend not only on the absolute value of the impurity
position, but also on the direction of the shift from the center
of the wire. This is because in general the noise depends on the
measurement point $x$, which is located in one of the leads, and
therefore a shift of the impurity away from the center implies a
smaller or bigger distance from the measurement point $x$.

The spikes in the spectrum of $S_{\rm imp}$  are essentially due
to resonances caused by the Andreev-type reflections at the
contacts and backscattering at the impurity. A difference with
respect to the clean case should be emphasized: In the clean case
the peaks of the conductivity $\sigma_0(x,x,\omega)$ are due to
resonant processes where a current pulse propagates from the
measurement point $x$ and comes back to~$x$ after a sequence of
Andreev reflections at the contacts. In contrast, in the presence
of the impurity, the conductivity $\sigma(x,x,\omega)$ also
includes processes, where a current pulse from the measurement
point $x$ propagates to the impurity position $x_0$, and then back
from $x_0$ to the measurement point $x$, whereby each path segment
connecting $x$ and $x_0$ includes a sequence of Andreev-type
reflections. These processes are described by $\sigma_{\rm BS}$
given in Eq.~(\ref{sigbs}), which for $x=y$ contains the product
of $\sigma_0(x,x_0,\omega)$ and $\sigma_0(x_0,x,\omega)$. This
product is weighted by a time integral over the exponential of the
Bose field correlation function at the impurity position, which in
turn strongly depends on the interaction strength. As a
consequence of this complicated interference of many different
contributions, the analytical determination of the location of the
peaks of $S_{\rm imp}$ is a complex problem, in which four
elementary frequencies combine: $\omega_{x}= v_F/(|x|-L/2$),
related to the distance of the measurement point from the closest
contact, $\omega_L$, the ballistic frequency of the finite length
wire, and $\omega_{\pm}=v_F/g(1/2 \pm \xi_0)$ related to the
distances of the impurity from the two contacts. A straightforward
analytic formula for the peak positions is thus in general not
available, however, progress can be made in some
 limits. For strong interaction $(g \lesssim 0.4)$ and in
the range of moderate to high frequencies ($\omega \gtrsim 3
\omega_L$), the spectrum of $S_{\rm imp}$ is essentially
determined   by   $\Im [\sigma^2_0(x,x_0,\omega)]$. An analysis of
this quantity shows that peaks occur at frequencies $\omega= n \pi
\omega_L$ ($n$ integer) whenever there is an integer $m$ such that
$2 n (\omega_{x}^{-1} \pm \Omega^{-1} ) \simeq
\omega_L^{-1}(1/2+m)$, where $\Omega$ can be either $\omega_L$,
$\omega_{+}$, $\omega_{-}$ or their sums and differences. The
peaks are positive (negative) if $m$ is even (odd). Thus, one can
realize that for some values of the impurity position, an upward
spike due to a resonance with a frequency $\Omega$ can be located
close to a downward spike related to another frequency
$\Omega^{'}$, and therefore abrupt changes of the sign of  $S_{\rm
imp}$ may occur, like in the solid curve of Fig.~\ref{noise_eq_2}
near $\omega= 2 \pi \omega_L$ and $\omega= 8 \pi \omega_L$.

On the other hand, for very weak interaction ($g \gtrsim 0.85$)
the time-integral in Eq.~(\ref{sigbs}) over the exponential of the
Bose field correlation function grows as a power law $|\omega|^{2
g-1}$, and therefore plays an important role. At low frequencies
$0 < \omega/\omega_L < 10$ one finds that $S_{\rm imp}$ is
essentially proportional to $ |\omega|^{2 g-1} \,
(\Re[\sigma_0(x,x_0,\omega)])^2$, which roughly amounts to $S_{\rm
imp} \sim |\omega|^{2 g-1} \cos^2 [\omega
(\delta_x/g-1/2+\xi_0)/\omega_L] $. The frequency range in which
this approximation is valid is however also interaction dependent.
In particular, the approximation becomes valid for all frequencies
when the interaction is switched off ($g \rightarrow 1$). The
different forms of the frequency dependence of $S_{\rm imp}$ for
various values of the interaction strength $g$ are shown in
Fig.~\ref{noise_eq_3}, where the limiting cases discussed above
can be recognized.

%%%%%%%%%%%%%%%%%%%%%%%%%%%%%%%%%%%%%%%%%%%%%%%%%
%%%%%%%%%%%%%%%%%%%%%%%%%%%%%%%%%%%%%%%%%%%%%%%%%
%%%%%%%%%%%%%%%%%%%%%%%%%%%%%%%%%%%%%%%%%%%%%%%%%
\subsection{Non equilibrium noise}
\label{sec_non_eq}

Usually the FDT (\ref{s0thermal}) is only valid in thermal
equilibrium. However, in the absence of an impurity, the current
operator of a QW attached to Fermi liquid leads is a linear
superposition of contributions from normal plasmon modes. Since
the only mode depending on the voltage (i.e. the zero mode) is
noiseless, one can in fact conclude that $S_0(x,x,\omega)$ does
not depend on the applied voltage~$V$ and is just given by
Eq.~(\ref{s0thermal}), also out of equilibrium. In contrast, the
impurity noise~$S_{\rm imp}(x,x,\omega)$ is voltage dependent.
Again we concentrate on the weak backscattering limit which
amounts to performing a perturbative expansion in $\lambda$, up to
order $\lambda^2$. In doing so the average value $\langle \ldots
\rangle_{\rightarrow}$ can actually be replaced by the free
average $\langle \ldots \rangle_{0}$. Thus, it can be shown that
(up to order $\lambda^2$) the two contributions $S_A$ and $S_C$ to
the impurity noise (\ref{ff_result}) read
\begin{widetext}
\begin{eqnarray}
S_A(x,x,\omega) &=& \frac{1}{4 \pi} \left(\frac{h}{e^2}\right)^2
|\sigma_0(x,x_0,\omega)|^2 f_A^{(2)}(x_0,\omega) \; ,
\label{S_A_lambda2} \\
S_C(x,x,\omega) &=& -\frac{1}{\pi} \left(\frac{h}{e^2} \right)^2
\coth\left( \frac{\hbar \omega}{2 k_B T} \right) \Re
[\sigma_0(x,x_0,\omega)] \, \Re \left[  \sigma_0(x,x_0,\omega)
 f_C^{(2)}(x_0,\omega)   \right]  \label{S_C_lambda2}
\end{eqnarray}
with
\begin{eqnarray}
f_A^{(2)}(x_0,\omega) &=&  4 \pi \left( \frac{e \lambda}{\hbar}
\right)^2 \int_{0}^\infty dt \cos(\omega t) \cos(\omega_0 t)
\sum_{s=\pm} e^{4 \pi {C}_0(x_0,st;x_0,0)}  \; %\label{s_A_lambda2}
\nonumber \\
 &=&  4 \pi \left( \frac{e
\lambda}{\hbar} \right)^2 \frac{i}{2} \sum_{ q=\pm} \coth\left(
\frac{ \hbar(\omega + q \omega_0)}{k_B T} \right) \int_{0}^\infty
dt \, \sin((\omega + q \omega_0) t) \sum_{s=\pm} s e^{4 \pi
{C}_0(x_0,st;x_0,0)}  \; , \label{s_A_lambda2_bis}
\end{eqnarray}
and
\begin{eqnarray}
f_C^{(2)}(x_0,\omega) &=& 2\pi \left( \frac{e \lambda}{\hbar}
\right)^2 \int_0^\infty dt \, (e^{i  \omega t}-1) \cos(\omega_0 t)
\, \sum_{s=\pm} s e^{4\pi C_0 (x_0,st;x_0,0)} \; ,
\label{s_C_lambda2}
\end{eqnarray}
where $\omega_0=eV/\hbar$, and
\begin{equation}
\sigma_0(x,x_0,\omega)= (1-\gamma) \frac{ e^2}{h}    \frac{ e^{i
\omega (\frac{x}{L}-\frac{1}{2})/g \omega_L}}{1-\gamma^{2} e^{2 i
\omega/\omega_L }} \left( e^{i \frac{\omega}{\omega_L}
(\frac{1}{2}-\xi_0) }   + \gamma  e^{i
\frac{\omega}{\omega_L}(\frac{3}{2}+\xi_0)}
  \right) \; . \label{sigma0-wire-lead}
\end{equation}
\end{widetext}
In Eq.~(\ref{sigma0-wire-lead}), $x$ is assumed to be in the right
lead. Importantly, the modulus square of
Eq.~(\ref{sigma0-wire-lead})
\begin{eqnarray} \label{modofsigma}
\lefteqn{\left(\frac{h}{e^2}\right)^2 |\sigma_0(x,x_0,\omega)|^2 =
}
& &\nonumber \\
&& (1-\gamma)^2 \frac{1+\gamma^2+2\gamma\cos(2 \omega (\xi_0
+1/2)/\omega_L)}{1+\gamma^4-2\gamma^2\cos(2\omega/\omega_L)}
\end{eqnarray}
is independent of $x$, and thus $S_A(x,x,\omega)$ is independent
of $x$. However, the r.h.s.\ of Eq.~(\ref{S_A_lambda2}) depends on
the position of the impurity $x_0$. With the help of
Eq.~(\ref{s_A_lambda2_bis}) one sees that Eq.~(\ref{S_A_lambda2})
has a form similar to Eq.~(\ref{IBS}), the backscattering current
at the impurity.

%%%%%%%%%%%%%%%%%%%%%%%%%%%%%%%%%%%%%%%
%%%%%%%%%%%%%%%%%%%%%%%%%%%%%%%%%%%%%%%
%%%%%        FIGURE   12         %%%%%%
%%%%%%%%%%%%%%%%%%%%%%%%%%%%%%%%%%%%%%%
%%%%%%%%%%%%%%%%%%%%%%%%%%%%%%%%%%%%%%%
\begin{figure}
\begin{center}
\epsfig{file=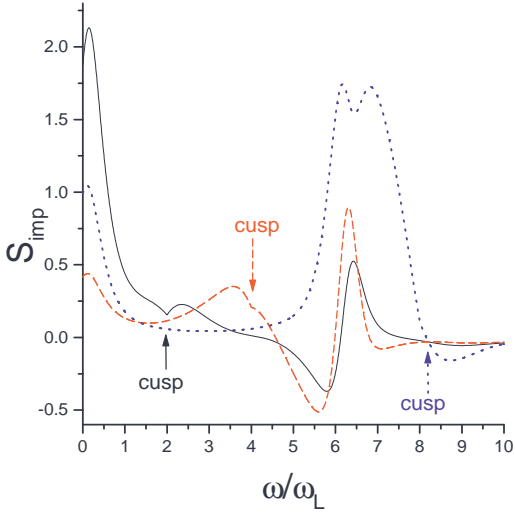,height=7cm,width=8cm}
\caption{\label{noise_neq_uVAR} The frequency spectrum of the
non-equilibrium impurity noise $S_{\rm imp}$ (in units of $e^2
\omega_L  (\lambda^{*}/\hbar \omega_L)^{2(1-g)}$), at $T=0$, for
three different values of the dimensionless voltage $u=eV/\hbar
\omega_L$: $u=2$ (solid curve), $u=4$ (dashed curve), and $u=8$
(dotted curve). The interaction strength is $g=0.25$ and the
impurity position is $\xi_0=0$. $S_{\rm imp}$ has a cusp
singularity at $\omega= eV/\hbar$. }
\end{center}
\end{figure}

%%%%%%%%%%%%%%%%%%%%%%%%%%%%%%%%%%%%%%%
%%%%%%%%%%%%%%%%%%%%%%%%%%%%%%%%%%%%%%%
%%%%%        FIGURE   13         %%%%%%
%%%%%%%%%%%%%%%%%%%%%%%%%%%%%%%%%%%%%%%
%%%%%%%%%%%%%%%%%%%%%%%%%%%%%%%%%%%%%%%
\begin{figure}
\begin{center}
\epsfig{file=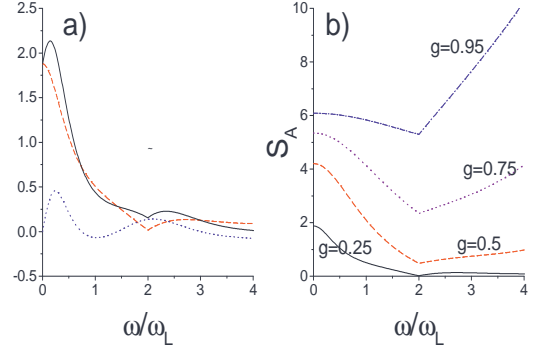,height=5.5cm,width=8cm}
\caption{\label{noise_neq_SVAR} a) The frequency spectra of the
two contributions $S_A$ (dashed curve) and $S_C$ (dotted curve) to
the impurity noise $S_{\rm imp}=S_A+S_C$ (solid curve) show that
the presence of the cusp is due to $S_A$. The value of the
dimensionless voltage is $u=2$, and the other parameters are
chosen as in Fig.~\ref{noise_neq_uVAR}.
\\b) The contribution $S_A$ for four different values of the
interaction strength: $g=0.25$ (solid curve), $g=0.50$ (dashed
curve), $g=0.75$ (dotted curve), and $g=0.95$ (dash-dotted curve).
The interaction modifies the slope to the right and to the left of
the singularity at $\omega= eV/ \hbar$.}
\end{center}
\end{figure}
We discuss now the non-equilibrium properties of the spectrum
$S_{\rm imp}(x,x,\omega)$ of the impurity noise, considering for
simplicity the case of a centered impurity at $\xi_0=0$. We first
analyze the case of zero temperature, where  a non-analyticity at
the frequency $\omega_0=eV/\hbar$ arises. In
Fig.~\ref{noise_neq_uVAR}, one can see the discontinuity of the
derivative of $S_{\rm imp}$ with respect to the frequency. This
singularity stems from the contribution $S_A$ to the impurity
noise $S_{\rm imp}$, as shown in Fig.~\ref{noise_neq_SVAR}a. Thus,
in regimes where $S_A$ is dominated by $S_C$, the singularity is
hardly visible: this is the case when the interaction is strong
and when $\omega \sim \omega_0 \gg \omega_L$, as for the dotted
curve in Fig.~\ref{noise_neq_uVAR}.

This non-analyticity of the noise at $\omega=\pm \omega_0$ is
already present in a non-interacting wire where it arises from the
sharpness of the Fermi surface at $T=0$. \cite{lesov97} This is
recovered in our model by taking the limit $g \rightarrow 1$. In
this limit, $S_A$ turns out to describe current fluctuations due
to electrons originating from different reservoirs (the left and
the right one), whereas $S_C$ comes from current fluctuations due
to electrons originating from the same reservoir (see
e.g.~Ref.~[\onlinecite{blanter00}]). When electron-electron
interaction is taken into account ($g<1$), the slope of the
branches of $S_A$ below and above the singularity is modified.
While for a non-interacting system $S_A$ is frequency-independent
for $ \omega \lesssim \omega_0$ and proportional to
$\omega-\omega_0$ for $ \omega \gtrsim \omega_0$,
in the interacting case, one finds interaction dependent
slopes, clearly shown  in Fig.~\ref{noise_neq_SVAR}b. Thus, the
type of singularity at $\omega= \pm \omega_0$  turns out to be
essentially unchanged by the interaction, in particular, the
location of the cusp is not modified. We emphasize that this
prediction is different from the results for  the homogeneous TLL,
where the linear dependence $\propto |\omega \pm \omega_0|$ is
turned into   a $g$-dependent power-law $| \omega \pm
\omega_0|^{2g-1}$. \cite{chamo96,chamo99} Hence, this power-law is
masked by the presence of the leads.

In the case of finite temperature, the cusp at $\omega=eV/\hbar$
is smeared like in a non-interacting wire, i.e.\ through a
$\coth[\hbar (\omega \pm \omega_0)/k_B T]$ prefactor (see
Eq.~(\ref{s_A_lambda2_bis})).

%%%%%%%%%%%%%%%%%%%%%%%%%%%%%%%
%%%%%%%%%%%%%%%%%%%%%%%%%%%%%%%
%%%%%%%%%%%%%%%%%%%%%%%%%%%%%%%
%%%%%%%%%%%%%%%%%%%%%%%%%%%%%%%
%%%%%%%%%%%%%%%%%%%%%%%%%%%%%%%
%%%%    SHOT NOISE LIMIT   %%%%
%%%%%%%%%%%%%%%%%%%%%%%%%%%%%%%
%%%%%%%%%%%%%%%%%%%%%%%%%%%%%%%
%%%%%%%%%%%%%%%%%%%%%%%%%%%%%%%
%%%%%%%%%%%%%%%%%%%%%%%%%%%%%%%
%%%%%%%%%%%%%%%%%%%%%%%%%%%%%%%

\subsubsection{Shot noise regime}
\label{sec_sn}

The ideal shot noise is defined as the value of the noise at zero
temperature and zero frequency. In this limit, the ratio of the
current noise and the backscattering current is related to the
charge that is transferred in each backscattering event.
\cite{blanter00} For the system under consideration, it has
already been shown \cite{ponomarenko_sn,trauz_sn} that the shot
noise is independent of $x$ and can be written as
\begin{equation} \label{pureshotnoise}
S(x,x,0) = 2 e I_{\rm BS} \; ,
\end{equation}
where the backscattering current $I_{\rm BS}$ is given by
Eq.~(\ref{IBS_nonpert}). This result can be easily seen from the
expressions for the current and noise at order $\lambda^2$ derived
above. For $T=0$ and $\omega=0$ the terms in Eqs.~(\ref{IBS}) and
(\ref{s_A_lambda2_bis}) with '$s=-1$' vanish and $S_A(x,x,0)$ is
the only contribution to the noise ($S_0$ and $S_C$ vanish for
$T=0$ and $\omega=0$). This is why $S(x,x,0)$ and $I_{\rm BS}$ are
related to each other by the simple relation
(\ref{pureshotnoise}). Hence, if $\omega \ll \omega_L$, the
fractional charge of the charge excitations that are backscattered
at an impurity in a TLL \cite{fisherGlazman,pham00} does not
become visible through the Fano factor $F = S/2 e I_{\rm BS}$.
Instead, Eq.~(\ref{pureshotnoise}) just reveals the electron
charge $e$.

Since this is a consequence of the fact that at very low
frequencies one probes the charge dynamics at long length scales,
larger than the wire length $L$,  it should be interesting to look
at frequencies near $\omega_L$, where the current noise is
expected to become sensitive to internal backscattering processes
in the wire. In particular, it will be shown  below that a voltage
and temperature regime exists in which the frequency behavior of
the noise is given by
\begin{equation} \label{shot-IBS}
S(x,x,\omega) \simeq 2 e F(\omega) I_{\rm BS} \; .
\end{equation}
Here the function $F(\omega)$, which plays the role of an
effective Fano factor, reads
\begin{equation} \label{fanofac}
F(\omega) = \frac{h^2}{e^4} \left| \sigma_0(x,x_0,\omega)
\right|^2 \; ,
\end{equation}
where $\sigma_0(x,x_0,\omega)$ is given by Eq.~(\ref{modofsigma}).
\ Before discussing in detail the parameter regime in which the
noise acquires the simple form (\ref{shot-IBS}), we want to
comment on the physical contents of this relation.

As  shown in Ref.~[\onlinecite{trauz04}], the effective Fano
factor  $F(\omega)$ allows to determine the fractional charge of
the quasiparticles of the interacting wire. Importantly,
Eq.~(\ref{fanofac}) does not depend on temperature since the
conductivity of the clean wire is temperature independent. An
explicit expression for $F(\omega)$ is obtained by inserting
Eq.~(\ref{modofsigma}) into Eq.~(\ref{fanofac}). In doing so, one
can realize that $F$ is actually also independent of the point of
measurement $x$, but it depends on the impurity position $x_0$.

\begin{figure}
\begin{center}
\epsfig{file=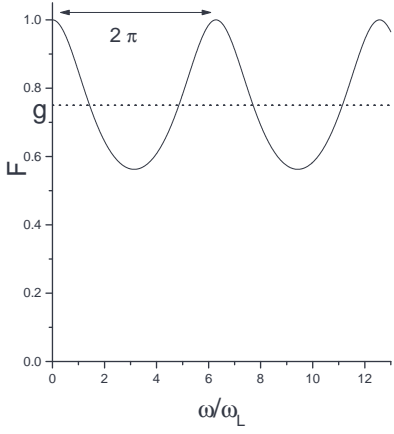,height=6cm,width=6cm}
\caption{\label{Fano} The periodic function $F(\omega)$ defined in
Eq.~(\ref{fofomega}) is plotted for $g=0.75$ against
$\omega/\omega_L$. While $F(0)=1$,  the average over one period
yields the interaction parameter $g$.}
\end{center}
\end{figure}

We first describe the case of a centered impurity, i.e.\
$\xi_0=x_0/L=0$; in this case the Fano factor reads
\begin{equation} \label{fofomega}
F (\omega) = \frac{2g^2}{1+g^2-(1-g^2) \cos (\omega/\omega_L)} \, ,
\end{equation}
and its behavior as a function of $\omega$ is shown in
Fig.~\ref{Fano}. The choice of a centered impurity also allows us
to compare the noise of the finite length wire with the noise
obtained within the homogeneous TLL model. The latter model
corresponds, in a sense, to the limit $L \rightarrow \infty$ of
our model with a fixed impurity position $x_0$, hence,
$\xi_0=x_0/L \rightarrow 0$.

Let us discuss several limits of Eq.~(\ref{shot-IBS}). Since
$F(0)=1$, in the limit $\omega \rightarrow 0$, Eq.~(\ref{shot-IBS})
yields the interaction-independent result (\ref{pureshotnoise}).
Evidently, one has to go to finite frequency to obtain more
interesting results. If we again adopt the averaging procedure
(\ref{average-fast}), and note that the average value of
$F(\omega)$ is the interaction parameter $g$ (see
Fig.~\ref{Fano}), we obtain\cite{NOTA2}
\begin{equation}
\left\langle S(x,x,\omega) \right\rangle_{\Delta \omega_1} \simeq
2 e g I_{BS} \, . \label{s_ex_average}
\end{equation}
Seemingly, Eq.~(\ref{s_ex_average}) suggests that quasiparticles
with a fractional charge $e^*=eg$ are backscattered off the
impurity in the TLL. For $L \rightarrow \infty$ the Fano factor
(\ref{fofomega}) becomes a rapidly oscillating function, and the
frequency interval $\Delta \omega_1$ over which $F(\omega)$ is
averaged in Eq.~(\ref{s_ex_average}) vanishes. This result should
therefore describe the noise of the homogeneous TLL model with an
impurity \cite{kane_fisher_noise} as it is indeed the case. Hence,
our findings are in accordance with previous considerations of the
effects of an impurity in a TLL. \cite{fisherGlazman,pham00}

We now discuss  in detail the conditions  under which the simple
expression (\ref{shot-IBS}) holds. In particular, we consider the
parameter regime $eV \gg \{ k_B T, \hbar \omega, \hbar \omega_L
\}$ (which will be referred henceforth as the {\it shot noise
regime}). We write the noise as
\begin{equation}
S(x,x,\omega) = 2 e F(\omega) \, I_{\rm BS} \, \left[ \Delta_A \,
 + \Delta_C  + \, \Delta_0 \right]
\end{equation}
with the functions $\Delta_\alpha={S_\alpha}/ 2 e F(\omega) I_{\rm
BS}$, where the subscript $\alpha$ takes the values
$\alpha=A,C,0$, and the $S_\alpha$ are respectively given by
Eqs.~(\ref{s0thermal}), (\ref{S_A_lambda2}), and
(\ref{S_C_lambda2}). Using these formulas, one can then evaluate
$\Delta_A$, $\Delta_C$, and $\Delta_0$ in the shot noise regime.
The range of validity of Eq.~(\ref{shot-IBS}) corresponds to a
parameter regime in which $\Delta_A \simeq 1$, $|\Delta_C|,
|\Delta_0| \ll 1$. The main contribution to the shot noise thus
stems from $S_A$.

\begin{figure}
\begin{center}
\epsfig{file=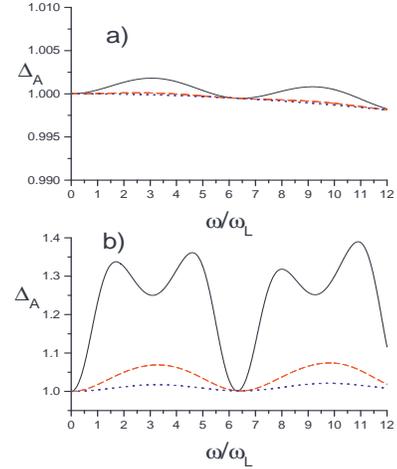,height=7cm,width=6cm}
\caption{\label{Fig-DeltaA} The dimensionless quantity $\Delta_A$
defined in the text as a function of frequency in the shot noise
regime $eV \gg \{ k_B T, \hbar \omega, \hbar \omega_L \}$. The
value of the dimensionless voltage is $u=eV/\hbar\omega_L=100$,
and the impurity is located in the middle $\xi_0=0$. Differences
between the exact value of $\Delta_A$ and its approximation
(\ref{DeltaA}) are not visible. a) refers to the case of weak
interaction $g=0.75$ and b) to the case of strong interaction
$g=0.25$. The three different curves in each plot correspond to
three different values of the dimensionless temperatures
$\Theta=k_B T/\hbar \omega_L$, $\Theta=0$ (solid), $\Theta=1$
(dashed), and $\Theta=2$ (dotted). Notice the different scales in
the two cases: For weak interaction even small temperatures allow
to fulfill the requirement $\Delta_A \simeq 1$, whereas for strong
interaction a temperature of the order of $\Theta \simeq 2-3$ is
required. }
\end{center}
\end{figure}

Indeed, in the shot noise regime $S_A$ can easily be shown to read
\begin{eqnarray}\nonumber
S_A(x,x,\omega) &=&  e \, F(\omega)  \left[ I_{\rm BS}(eV+\hbar
\omega) + I_{\rm BS}(eV-\hbar \omega)
\right]\\
&\times& \left\{ 1 + \mathcal{O} \left[ \exp{\left(\frac{k_B T}{eV
\pm \hbar \omega}\right)} \right] \right\} \label{SAshot}
\end{eqnarray}
which gives
\begin{equation}
 \Delta_A  \, \simeq \,  \frac{I_{\rm BS}(eV +  \hbar
\omega) +I_{\rm BS}(eV -  \hbar \omega) }{2 I_{\rm BS}(eV)} \, \,
\, , \label{DeltaA}
\end{equation}
where the omitted terms are exponentially small corrections.
Eq.~(\ref{DeltaA}) reveals that $\Delta_A$ can be directly
extracted from the current voltage characteristics. As shown in
Fig.~\ref{Fig-DeltaA} for the case of $eV=100 \hbar \omega_L$, in
order to make the deviations from $\Delta_A = 1$ negligible, one
has to operate at a sufficiently high temperature, so that the
oscillations of the current are damped. For weak interactions ($g
\simeq 0.75$) a temperature of the order of $k_B T \simeq \hbar
\omega_L$ or even smaller is already sufficient to fulfill this
requirement, whereas for strong interaction ($g\simeq 0.25$)
higher temperatures of the order of $k_B T \simeq 2-3 \hbar
\omega_L$ are necessary.
%%%%%%%%%%%%%%%%%%%%%%%%%%%%%%%%%%%%%%%
%%%%%%%%%%%%%%%%%%%%%%%%%%%%%%%%%%%%%%%
%%%%%      FIGURE    14          %%%%%%
%%%%%%%%%%%%%%%%%%%%%%%%%%%%%%%%%%%%%%%
%%%%%%%%%%%%%%%%%%%%%%%%%%%%%%%%%%%%%%%
%%%%%%%%%%%%%%%%%%%%%%%%%%%%%%%%%%%%%%%
%%%%%%%%%%%%%%%%%%%%%%%%%%%%%%%%%%%%%%%
%%%%%      FIGURE    15          %%%%%%
%%%%%%%%%%%%%%%%%%%%%%%%%%%%%%%%%%%%%%%
%%%%%%%%%%%%%%%%%%%%%%%%%%%%%%%%%%%%%%%

We now turn to the correction $\Delta_C$. In the low frequencies
range ($\omega \lesssim \omega_L$), $\Delta_C$ can be written as
\begin{equation}
|\Delta_C| = \frac{2 \hbar \omega}{ eV} \coth \left( \frac{\hbar
\omega}{2 k_B T} \right) \, \left[ |\kappa_{\rm BS}(V)|  +
\mathcal{O}  \left( \frac{\omega}{\omega_L} \right)^2 \right] \, ,
\label{DeltaC}
\end{equation}
where  the quantity $\kappa_{\rm BS}(V)$ is defined in
Eq.~(\ref{kappa_def}). Here, one can again distinguish two cases.
For weak interaction, $\kappa_{\rm BS}(V)$ is a slowly decreasing
function of voltage, of the order of $2g-1$ (see
Fig.~\ref{figM1_1}b). Therefore $|\Delta_C| \ll 1$  requires $eV
\gg \max \{ k_B T, \hbar \omega , \hbar \omega_L \}$. In contrast,
for strong interaction $\kappa_{\rm BS}(V)$ exhibits large
oscillations increasing with voltage (see Fig.~\ref{figM1_1}a),
which might yield very large values of $\Delta_C$. To recover the
simple behavior (\ref{shot-IBS}), it is therefore crucial to
exploit these oscillations by choosing voltage values at which
$\kappa_{\rm BS}(V) \thickapprox 0$. Thus, with a suitable choice
of temperature and voltage, the condition $|\Delta_C| \ll 1$ can
always be fulfilled in the low frequency range. When this is the
case, it turns out from the numerical analysis that the condition
$|\Delta_C| \ll 1$ then remains valid up to frequencies $\omega
\sim 10 \, \omega_L$.

Finally, to obtain an estimate of $\Delta_0$, we observe that the
backscattering current can be written as $I_{\rm BS} = (e^2/h)
\mathcal{R} V$, where $\mathcal{R} \ll 1$ is an effective
reflection coefficient that can be directly extracted from a
current-voltage measurement. In contrast to a non-interacting
wire, $\mathcal{R}$ now depends in general on voltage, interaction
strength, and temperature. Since the (temperature independent)
function $(h/e^2)\Re[\sigma_0(x,x,\omega)]/F(\omega)$ has a
maximum ${\cal C}$, we obtain for $\Delta_0$  the  upper bound
\begin{equation}
|\Delta_0| \le \coth \left( \frac{\hbar \omega}{2 k_B T} \right)
\frac{ \hbar \omega}{ eV \mathcal{R}} {\cal C} \, . \label{Delta0}
\end{equation}
The value of ${\cal C}$ depends on the interaction strength. For
relatively weak interaction ($g \simeq 0.75$) one has ${\cal C}
\simeq 1$, whereas for strong interaction ($g \simeq 0.25$) ${\cal
C} \simeq 4$. Thus, the condition $|\Delta_0| \ll 1$ is certainly
fulfilled in the parameter range $eV \mathcal{R} \gg {\cal C} \max
\{ k_B T, \hbar \omega , \hbar \omega_L \}$.
%%%%%%%%%%%%%%%%%%%%%%%%%%%%%%%%%%%%%%%
%%%%%%%%%%%%%%%%%%%%%%%%%%%%%%%%%%%%%%%
%%%%%        FIGURE   16         %%%%%%
%%%%%%%%%%%%%%%%%%%%%%%%%%%%%%%%%%%%%%%
%%%%%%%%%%%%%%%%%%%%%%%%%%%%%%%%%%%%%%%
\begin{figure}
\begin{center}
\epsfig{file=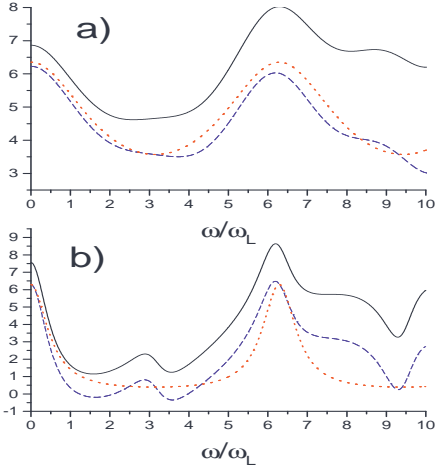,height=7cm,width=7cm}
\caption{\label{full_noise1} The frequency spectrum of the full
noise $S(x,x,\omega)$ (solid line) in units of $e^2 \omega_L$ and
the approximation (\ref{shot-IBS}) (dotted line) are depicted for
two systems with the parameters a) $g=0.75$, $\Theta=1$, $u=100$,
$\delta_x=0.05$, $\xi_0=0$, $\mathcal{R}=0.2$ and b) $g=0.25$,
$\Theta=2$, $u=98.83$, $\delta_x=0.05$, $\xi_0=0$,
$\mathcal{R}=0.2$. The dashed curve is obtained after a background
subtraction (see text).}
\end{center}
\end{figure}

In summarizing these considerations, we see that a noise
measurement in the parameter regime $eV \mathcal{R} \gg {\cal C}
\hbar \omega$ allows to extract the Fano factor $F(\omega)$ from
noise data. The optimal temperature at which one should operate
depends on the QW under investigation. In particular, for systems
characterized by weak or moderate interaction ($g \simeq 0.75$),
like for most semiconductor QWs, the optimal temperature is $k_B
T/\hbar \omega_L \lesssim 1$, while for strong interaction ($g
\simeq 0.25$), like for SWNTs, a temperature $k_B T / \hbar
\omega_L \simeq 2-3 $ is more appropriate. On the other hand, the
voltage needs to be rather high to fulfill in particular the
condition $\Delta_0\ll 1$. However, this problem can be at least
partially overcome by an appropriate background subtraction. This
is demonstrated in Fig.~\ref{full_noise1} for the cases of weak
and strong interaction, respectively.

Fig.~\ref{full_noise1}a refers to a weakly interacting QW with
interaction parameter $g=0.75$. The temperature
$k_BT=\hbar\omega_L$ and the voltage $eV=100\hbar\omega_L$. The
impurity is assumed to be in the middle, i.e., $\xi_0=0$, and the
measurement point is near a contact so that $\delta_x=0.05$. The
impurity leads to an effective reflection coefficient
$\mathcal{R}=0.2$ which correspond to an effective impurity
strength $\lambda^*\sim 10^{-4}eV$. For these parameters the full
noise has still a sizable contribution from $S_0$, as can be
estimated from Eq.~(\ref{Delta0}), which gives for frequencies
$\omega\sim \pi\omega_L$ and the above parameters $\Delta_0\sim
0.2$. A simplified estimate for $S_0$ reads
\begin{equation}\nonumber
S_{\rm off-set} = \frac{e^2}{\pi} \omega \coth( \hbar \omega/2 k_B
T)\, ,
\end{equation}
which can be subtracted from the full noise to give the dashed
curve in Fig.~\ref{full_noise1}a. When this curve is averaged
between $\omega=0$ and the first maximum, one extracts $g\sim
0.71$, which is a reasonable estimate of 0.75.

Fig.~\ref{full_noise1}b refers to a strongly interacting QW with
interaction parameter $g=0.25$. Here, the temperature
$k_BT=2\hbar\omega_L$ and the voltage $eV=98.83\hbar\omega_L$, a
value chosen with the help of the current voltage characteristic
in such a way that $\kappa_{\rm BS}=0$. This makes the
contribution of $S_C$ negligible, as discussed above, cf.\
Eq.~(\ref{DeltaC}). The impurity is again assumed to be in the
middle, i.e., $\xi_0=0$, and the measurement point corresponds to
$\delta_x=0.05$. The effective reflection coefficient
$\mathcal{R}=0.2$ now comes from an impurity with effective
strength $\lambda^*\sim 0.12 eV$. For these parameters the full
noise has a large contribution from $S_0$, since
Eq.~(\ref{Delta0}) gives $\Delta_0\sim 0.6$ for frequencies
$\omega\sim \pi\omega_L$. Despite this large value of $\Delta_0$,
the value of $g$ can still be extracted. A subtraction of $S_{\rm
off-set}$ leads to the dashed curve in Fig.~\ref{full_noise1}b.
When this curve is averaged between $\omega=0$ and the second
maximum, where the noise reaches again the same level as at low
frequencies, one obtains $g\sim 0.23$, again a reasonable estimate
of 0.25. A more accurate data analysis can, of course, be based on
the full expressions for the noise derived above.

We mention that a similar procedure can be provided for the case
of an off-centered impurity. For an arbitrary impurity position
$x_0$, the Fano factor reads
\begin{eqnarray} \label{Fano-fun}
F(\omega) =   ( 1-\gamma)^2 \frac{1+\gamma^2+2\gamma \cos \left(
\frac{2\omega (\xi_0+1/2)}{\omega_L} \right)}{1+\gamma^4-2\gamma^2
\cos \left( \frac{2 \omega}{\omega_L}\right)} \, ,
\end{eqnarray}
where again $\xi_0=x_0/L$. When compared to the case of a centered
impurity, we observe that now for certain noise frequencies a
pronounced reduction of the Fano factor occurs, as discussed in
Ref.~[\onlinecite{trauz04}]. This effect could be related to the
low Fano factor measured in bundles of carbon
nanotubes.\cite{roche02} Secondly, since Eq.~(\ref{Fano-fun}) is
not in general a periodic function, the averaging procedure used
in Eq.~(\ref{s_ex_average}) to derive $g$ cannot be applied.
However, one can always introduce the function
\begin{equation}
Fi(\Omega) = \frac{1}{\Omega} \int_{0}^{\Omega} \, F(\omega) \,
d\omega \, ,
\end{equation}
which is a generalization of the averaging defined in
Eq.~(\ref{average-fast}). Interestingly, $Fi(\Omega)$ exhibits
oscillations around the value $g$ for any position $\xi_0$ of the
impurity; these oscillations are damped with increasing $\Omega$,
and the value of
$g$ can be extracted, or at least relatively well estimated.\\

Despite the progress made recently with the measurement of high
frequency noise, \cite{deblock,reulet} it might be difficult to
obtain accurate noise data up to frequencies of order
$10\omega_L$, if the QW is not very long. Then, a more detailed
analysis of low frequency noise data may be of particular
interest. As we have shown previously\cite{trauz04}, the
interaction constant $g$ could also be deduced from the
low-frequency behavior of the Fano factor (\ref{Fano-fun})
\[
F(\omega)=1-\left(\frac{\omega L}{2 v_F}\right)^2 \! \! (1-g^2)
\left[1+4 g^2 \xi_0(1+\xi_0)\right] +
\mathcal{O}\left(\frac{\omega}{\omega_L}\right)^4
\]
once the position of the impurity is known.  The latter can be
determined from the current voltage characteristics by tuning the
temperature, as described in Sec.~\ref{sec_cur}.\\

We conclude this discussion of the shot noise by mentioning
another interesting feature emerging in this regime, namely the
fact that the slope of the noise at $\omega=0$ is related to the
differential conductance.  Explicitly, we obtain
\begin{equation} \label{slope}
\lim_{\omega \rightarrow 0} \frac{S(x,x,\omega)-S(x,x,0)}{2 \hbar
\omega (e^2/h)} = \left( \frac{h}{e^2} \frac{dI}{dV} \right)^2 \;
.
\end{equation}
This result has been derived perturbatively in the impurity
strength including terms up to order $\lambda^2$ and is in
agreement with the Coulomb gas calculation for the homogeneous TLL
model. \cite{chamo99} Moreover, in the limit $g \rightarrow 1$,
Eq.~(\ref{slope}) coincides with the corresponding expression for
non-interacting electrons. \cite{blanter00} We note that the
limits $V \rightarrow 0$ and $\omega \rightarrow 0$ are not
interchangeable, since from Eq.~(\ref{FDT}) the slope of the noise
at equilibrium ($V=0$) is proportional to the linear conductance,
and not to its square. The simple relation Eq.~(\ref{slope})
between the slope of the noise and the dimensionless differential
conductance is only valid at zero temperature; in addition, an
exact solution of the homogeneous TLL model\cite{lesag97}
indicates that such a simple relation may not hold if all orders
in the impurity strength are taken into account in a
non-perturbative way.

%%%%%%%%%%%%%%%%%%%%%%%%%%%%%%%
%%%%    EXCESS NOISE       %%%%
%%%%%%%%%%%%%%%%%%%%%%%%%%%%%%%
\subsection{Excess noise}
\label{sec_exnoi}
%%%%%%%%%%%%%%%%%%%%%%%%%%%%%%%%%%%%%%%
%%%%%%%%%%%%%%%%%%%%%%%%%%%%%%%%%%%%%%%
%%%%%      FIGURE    17          %%%%%%
%%%%%%%%%%%%%%%%%%%%%%%%%%%%%%%%%%%%%%%
%%%%%%%%%%%%%%%%%%%%%%%%%%%%%%%%%%%%%%%
\begin{figure*}
\begin{center}
\epsfig{file=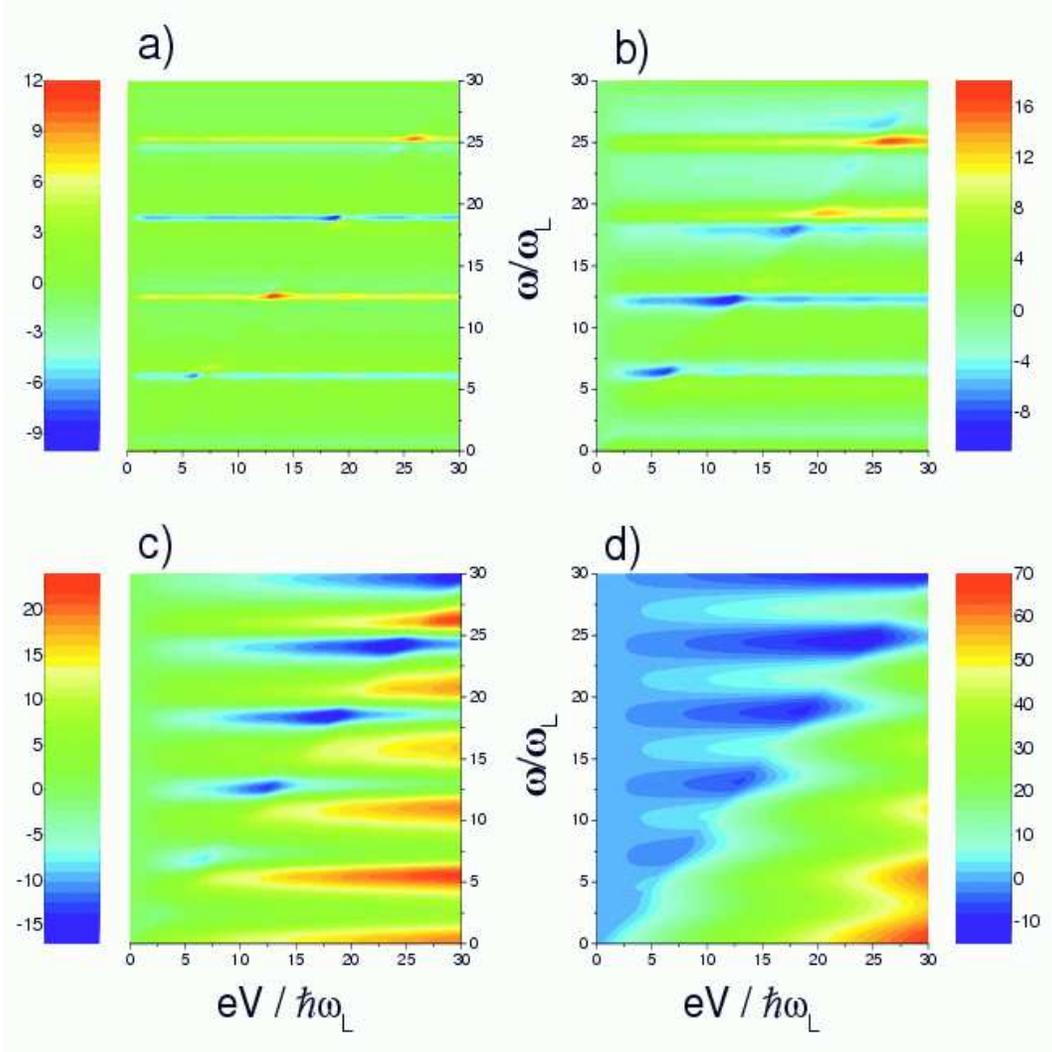,height=14cm,width=14cm,clip=}
\caption{\label{S_ex_plot} The excess noise at zero temperature
and for an impurity in the middle of the wire ($\xi_0=0$) is shown
(in units of $e^2 \omega_L  (\lambda^{*}/\hbar
\omega_L)^{2(1-g)}$) as a function of the dimensionless frequency
$\omega/\omega_L$ and the applied voltage $eV/\hbar \omega_L$ for
four values of interaction strength: a) $g=0.25$; b) $g=0.5$; c)
$g=0.75$; d) $g=0.95$. }
\end{center}
\end{figure*}
In the previous subsections we have discussed the equilibrium
noise and the non-equilibrium noise. In the present section we
focus on the excess noise, which is simply defined as their
difference
\begin{equation} \label{exnoise}
S_E(x,x,\omega) = S(x,x,\omega)-S(x,x,\omega)|_{V=0} \; .
\end{equation}
Since the $S_0$ contribution to the noise does not depend on the
voltage, $S_E$ is independent of $S_0$ and therefore a quantity
arising entirely from the presence of the impurity. For this
reason, the excess noise is of particular interest. The impurity
strength appears in $S_E$ as an overall scaling factor. In
contrast, in the full noise the presence of $S_0$, which is
independent of the impurity strength, requires a background
subtraction to reveal the interesting effects related to impurity
backscattering. We notice that $S_E$ can also be written as
\begin{equation} \label{exnoise2}
S_E(x,x,\omega) = S_{\rm imp}(x,x,\omega) \, - \, S_{\rm imp}
(x,x,\omega)|_{V=0}
\end{equation}
with
\[
S_{\rm imp} (x,x,\omega)|_{V=0} = 2\hbar \omega \coth \left(
\frac{\hbar \omega}{2k_BT} \right)  \Re [\sigma_{\rm
BS}(x,x,\omega)]  ,
\]
where $\sigma_{\rm BS}(x,y,\omega)$ is given in Eq.~(\ref{sigbs}).

The behavior of the excess noise is illustrated in
Fig.~\ref{S_ex_plot}. We first focus on Fig.~\ref{S_ex_plot}a,
which refers to the case of strong electron interaction
($g=0.25$). Then, apart from some special points along the
diagonal, the structure of the excess noise essentially exhibits a
voltage independent behavior, as the horizontally oriented shape
suggests. Indeed, for strong interaction and for moderate to high
frequencies ($\omega \gtrsim \omega_L$), the excess noise is
dominated by the equilibrium noise, in particular, by the
supplementary noise $S_{\rm imp}$ at equilibrium, i.e., $S_{E}
\simeq - S_{\rm imp} |_{V=0}$. Deviations from this behavior lead
to the spots along the diagonal of the excess noise diagram.
These spots appear whenever $eV/\hbar$ equals a frequency
corresponding to a peak of the equilibrium noise. Then, the excess
noise acquires a non negligible contribution also from the finite
voltage term, and the simple formula $S_{E} \simeq - S_{\rm imp}
|_{V=0}$ is not valid. As discussed in Sec.~\ref{sec-IV-A-2},
these particular frequency values are determined  by $\Im
[\sigma^2_0(x,x_0,\omega)]$.

Let us now consider the case of weaker interaction strength. The
evolution of the excess noise in Fig.~\ref{S_ex_plot} from a) to
d) shows that the scenario dramatically changes, even at the
qualitative level: The horizontal lines smear out and a diagonally
oriented shape arises. A very sharp diagonal structure is
precisely the expected behavior of the excess noise in the
noninteracting limit, where $S_E$ is given by\cite{blanter00}
\begin{equation}
S_E= \frac{e^2}{h}\left\{
\begin{array}{cc}
0 &   \hbar |\omega| > eV \\
& \\
2 (eV-\hbar|\omega|) \mathcal{R} (1-\mathcal{R}) &   \hbar
|\omega| < eV ,
\end{array}\right.
\end{equation}
where $\mathcal{R}$ is the reflection coefficient. Still, some
qualitative differences compared to the noninteracting case are
present. While for the non-interacting system the excess noise is
always strictly non-negative, regions of negative excess noise
emerge for the interacting case. The fact that the excess noise
$S_E$ can be {\it negative} may be surprising, since the presence
of an applied voltage is expected to increase the noise. However,
we deal here with a non-linear system, where the voltage also
affects the interference conditions between plasmonic charge
excitations that are Andreev-type reflected at the contacts and
backscattered by the impurity. Thus, depending on $V$, the
interference can be destructive or constructive, and -- under
certain circumstances -- the equilibrium noise can be larger than
the corresponding non-equilibrium noise at the same
noise frequency $\omega$.\\

%%%%%%%%%%%%%%%%%%%%%%%%%%%%%%%%%%%%%%%%%%%%%%%%%
%%%%%%%%%%%%%%%%%%%%%%%%%%%%%%%%%%%%%%%%%%%%%%%%%
%%%%%%%%%%%%%%%%%%%%%%%%%%%%%%%%%%%%%%%%%%%%%%%%%
%%%%%%       C O N C L U S I O N S        %%%%%%%
%%%%%%%%%%%%%%%%%%%%%%%%%%%%%%%%%%%%%%%%%%%%%%%%%
%%%%%%%%%%%%%%%%%%%%%%%%%%%%%%%%%%%%%%%%%%%%%%%%%
%%%%%%%%%%%%%%%%%%%%%%%%%%%%%%%%%%%%%%%%%%%%%%%%%

\section{Conclusions}
 \label{sec_con}

In the present work, we have investigated  transport properties of
interacting quantum wires with a weak impurity coupled to
non-interacting (Fermi liquid) electron reservoirs. We have used
the inhomogeneous Tomonaga Luttinger liquid model with a
backscattering term to describe the system, and the Keldysh
formalism to predict its transport properties out of equilibrium.
It has been shown that the finite length $L$ of the quantum wire
crucially affects the transport properties.

On the one hand, the nonlinear $I-V$ characteristics becomes an
oscillating function of the applied voltage, an effect which is
washed out if the temperature is raised. It has also been shown
that, with an appropriate choice of the temperature, these
oscillations can be exploited to measure both the interaction
strength~$g$  and the position of an impurity in the wire. We wish
to emphasize that these oscillations are not due to nonadiabatic
contacts, which would lead to ordinary Fabry-P{\'e}rot
oscillations that are also present in a noninteracting wire.
Instead, the oscillations predicted here arise because of a
mismatch of the strength of the electron-electron interaction in
the quantum wire and the leads. Therefore, a measurement of these
current oscillations would provide evidence for electronic
correlations in one-dimensional systems. From the above remarks it
is clear that one needs a way to distinguish experimentally the
current oscillations due to a mismatch of the electron-electron
interaction from ordinary Fabry-P{\'e}rot oscillations. We propose
to vary the voltage of a metallic back-gate that influences the
electron density in the quantum wire, since the ordinary
Fabry-P{\'e}rot oscillations are highly affected by a change of
the gate voltage. \cite{liang01,peca03} In contrast, this is not
necessarily the case for the oscillations predicted here. More
precisely, if the distance between the one-dimensional channel and
the gate electrode does not depend on the gate voltage, the
interaction parameter~$g$ is essentially independent of the
applied gate voltage, and the oscillations predicted here should
not be affected much by a change of the gate voltage. This will
typically be the case for a SWNT on an oxidized substrate. In
semiconductor QWs a weak dependence of $g$ on the gate voltage has
to be expected. However, the gate voltage dependence seems to be a
decisive factor distinguishing Fabry-P{\'e}rot oscillations from
the oscillations predicted by us.

On the other hand, we have shown that, similarly to the average
current, also the frequency spectrum of the current noise
significantly depends on the interaction strength and the finite
length of the wire. At frequencies $\omega \ll \omega_L = v_F/gL$,
the noise is determined by slow processes, much slower than the
traversal time of a plasmon through the interacting wire, and the
noise is not affected by the presence of electron-electron
interaction. This is similar to the absence of a suppression of
the DC conductance by the interaction in a one-dimensional quantum
wire connected to Fermi liquid leads. In the limit $\omega
\rightarrow 0$, internal properties of the interacting wire cannot
be resolved, and the wire with the impurity effectively behaves as
a complicated scatterer between two Fermi liquid leads. However,
for noise frequencies of the order of $\omega_L$, the situation
changes both for the equilibrium and the non-equilibrium case.

In particular the equilibrium noise, related to the AC
conductivity by the FDT, has been shown to exhibit oscillations as
a function of $\omega$ already in the absence of impurities in the
wire. These oscillations are characterized by two frequencies,
where the larger one is proportional to the ballistic frequency
related to the length of the wire, while the smaller one depends
on the distance of the measurement point of the noise from the
nearby contact. An average over the fast oscillations allows to
obtain a sinusoidal spectrum whose amplitude is directly connected
to the interaction strength. In the presence of an impurity there
is a supplementary equilibrium noise $S_{\rm imp}$. For a strongly
interacting QW, $S_{\rm imp}$ is characterized by sharp spikes due
to resonance phenomena caused by the Andreev-type reflections at
the contacts and the backscattering at the impurity.

In presence of an applied voltage, we have seen that the zero
temperature noise spectrum has cusps at the frequencies
$\omega=\pm eV/\hbar$, as it is already the case for a noninteracting
wire. We have shown that the electronic correlations affect the
value of the slopes on both sides of these singularities in a
different way. In contrast to predictions based on the homogeneous
TLL model, the type of singularity is however not changed.

We have then focussed on the shot noise regime ($eV \gg \{k_B T,
\hbar \omega, \hbar \omega_L \}$), where the system is strongly
out of equilibrium. In this case, the scattering processes at the
impurity are known to obey Poissonian statistics, and the noise is
proportional to the backscattering current. The value of the Fano
factor (ratio between $S$ and $I_{\rm BS}$) strongly depends on
the frequency range one explores. If $\omega$ is of the order of
the ballistic frequency $\omega_L$ or larger, the Fano factor
depends on $g$. This dependence becomes particularly simple if the
impurity is located in the middle of the wire. In that case,
$e^*=eg$ appears as a prefactor of the finite frequency noise,
when it is averaged over an appropriate frequency range. This may
allow for an independent way to measure $g$. The most promising
range of parameters for such an experiment as well as possible
improvements of the data analysis by a background subtractions
have been thoroughly discussed. Moreover, it has been shown that
relevant information is already contained  in the low frequency
noise data.

We also mention that in realistic one-dimensional systems, like
SWNTs or cleaved edge overgrowth QWs,  typical values of the Fermi
velocity are $v_F \sim 10^5 \div 10^6 {\rm m/s}$. The length of
the wires can vary from $1 \mu {\rm m}$ up to tenths of $\mu {\rm
m}$, whereas the typical interaction strengths are $g \sim 0.2$
for SWNTs and $g \sim 0.7$ for semiconductor QWs. These values
yield estimates of the ballistic frequency $\hbar \omega_L$ of
order meV or below.\cite{Lemay,Tserk} The conditions $eV \gg \hbar
\omega_L$ and $\omega \gg \omega_L$ for the observability of the
current oscillations and the fractional charge are therefore
experimentally realistic according to nowadays techniques.
\cite{deblock, schoelkopf} Moreover, the constraints for
experiments in the shot noise regime could be significantly
relaxed, provided that the experimental setup enables a
measurement of the supplementary noise $S_{\rm imp}=S_A+S_C$
instead of the full noise $S$ only. Then, the condition
$\Delta_0\ll 1$ is redundant, and one is left with the weaker
conditions $\Delta_A \simeq 1$ and $\Delta_C \ll 1$, which amount
to require $eV \gg \hbar \omega$, and $k_B T/\hbar \omega_L
\lesssim 1$ ($k_B T/\hbar \omega_L \sim 2-3$) for weak (strong)
interaction. Such experiments could be based on a quantum wire
with a tunable impurity, which could, for instance, be realized by
two crossed carbon nanotubes \cite{gao04} or a nanotube with a
nearby STM tip.

%%%%%%%%%%%%%%%%%%%%%%%%%%%%%%%%%%%%%%
%%%%%%%%%%%%%%%%%%%%%%%%%%%%%%%%%%%%%%
%%%%%%%%%%%%%%%%%%%%%%%%%%%%%%%%%%%%%%

\begin{acknowledgments}
Helpful discussions with H.~Bouchiat, R.~Deblock, R.~Egger,
D.~C.~Glattli, K.-V. Pham, F. Piechon, P.~Roche, and H. Saleur are
gratefully acknowledged. Financial support was provided by the EU
networks DIENOW and SPINTRONICS.\\
\end{acknowledgments}

\appendix
%%%%%%%%%%%%%%%%%%%%%%%%%%%%%%%%%%%%%%%%%%%%%%%%%
%%%%%%%%%%%%%%%%%%%%%%%%%%%%%%%%%%%%%%%%%%%%%%%%%
%%%%%%%%%%%%%%%%%%%%%%%%%%%%%%%%%%%%%%%%%%%%%%%%%
%%%%%%       A P P E N D I X    A         %%%%%%%
%%%%%%%%%%%%%%%%%%%%%%%%%%%%%%%%%%%%%%%%%%%%%%%%%
%%%%%%%%%%%%%%%%%%%%%%%%%%%%%%%%%%%%%%%%%%%%%%%%%
%%%%%%%%%%%%%%%%%%%%%%%%%%%%%%%%%%%%%%%%%%%%%%%%%

\section{Keldysh path-integral formulation}
\label{app_keld} In this appendix, we  provide  details of the
method to calculate  transport properties of the system by
adopting the Keldysh formalism. \cite{keldysh}

In the Hamiltonian (\ref{L}), the term (\ref{LV}) can equivalently
be replaced by
\begin{equation}
{\mathcal H}^{\prime}_V \, = \frac{1}{\sqrt{\pi}} \,
\int_{-\infty}^{+\infty}
\partial_x \mu(x) \, \Phi(x,t) dx \, . \label{Lprime-temp}
\end{equation}
Both forms  yield the same equations of motion; however, the
latter form is more suitable for the manipulations to be presented
below.

In the case of a constant applied voltage we have from
Eq.~(\ref{mu-profile})
\begin{equation}
\partial_x \mu = -\mu_L \delta(x+\frac{L}{2}) \, + \, \mu_R
\delta(x-\frac{L}{2}), \label{mu-deriv}
\end{equation}
but much of the following derivation is actually valid for any
function, including also time-dependent ones. For this reason,
instead of (\ref{Lprime-temp}), we shall consider here a general
term of the form
\begin{equation}
{\mathcal H}^{\prime}_V \, =   - \frac{e}{\sqrt{\pi}} \,
\int_{-\infty}^{+\infty} E(x,t) \, \Phi(x,t) dx \, . \label{LE}
\end{equation}

The reader will be alerted whenever the quantity $-e E(x,t)$ will
explicitly be replaced by the expression (\ref{mu-deriv}).

Let us now denote by $\Phi^+$ and $\Phi^-$ the complex fields on
the upper and lower time branch of the Keldysh contour and
introduce the generating functional
\begin{widetext}
\begin{eqnarray}
\displaystyle Z[J]&=& \frac{1}{\cal{N}_Z} \int {\mathcal D}
\Phi^{\pm} \, \exp \left\{-\frac{1}{2} \int d\mathbf{r'}
d\mathbf{r''} \sum_{\eta,\eta'=\pm} \Phi^{\eta}(\mathbf{r'})
({\mathcal C}^{-1}_0)^{\eta,\eta'}(\mathbf{r'},\mathbf{r''})
\Phi^{\eta'}(\mathbf{r''}) \right\} \nonumber
\\ & \times &  \exp{ \left\{\sum_{\eta=\pm} \left( -\frac{i}{\hbar} \eta
\int_{-\infty}^{+\infty} d t' \mathcal{H}_B[\Phi^{\eta}]+\frac{i e
}{\hbar \sqrt{\pi}}  \eta \int d\mathbf{r'}
E(\mathbf{r'})\Phi^\eta(\mathbf{r'}) +\frac{i}{\sqrt{2}} \int
d\mathbf{x} J(\mathbf{x})\Phi^\eta(\mathbf{x})\right)\right\}}\, ,
\label{gen-fun-1}
\end{eqnarray}
\end{widetext}
where  the vector label $\mathbf{r'}$ stands for
$\mathbf{r'}=(x',t')$, $\int d\mathbf{r'}=\int_{-\infty}^{+\infty}
dt' \int_{-\infty}^{+\infty} dx'$, and $\cal{N}_Z$ is a
normalization factor, which assures that $Z[0]=1$. In
Eq.~(\ref{gen-fun-1}), ${\mathcal
C}^{-1}_0(\mathbf{r'},\mathbf{r''})$ is the inverse of a $2 \times
2$ matrix defined by the four free correlators
\begin{equation}
{\mathcal C}^{\eta,\eta'}_0(\mathbf{r'};\mathbf{r''}) = \langle
\Phi^\eta(\mathbf{r'}) \Phi^{\eta'}(\mathbf{r''}) \rangle_0 \, ,
\end{equation}
where $\langle \ldots \rangle_0 $ indicates the average performed
with respect to the free Hamiltonian (\ref{L0}) along the Keldysh
contour.

For a wire with an impurity and in presence of an applied voltage,
we have
\begin{equation}
\langle \Phi(\mathbf{x}) \rangle = \frac{1}{2} \sum_{\eta=\pm}
\langle \Phi^\eta(\mathbf{x}) \rangle =\frac{-i}{\sqrt{2}} \left.
\frac{\delta Z[J]}{\delta J(\mathbf{x})} \right|_{J=0} \; .
\label{Phi-VM1}
\end{equation}
One can simplify the notation by introducing infinite-dimensional
vectors and matrices where {\it both} $\mathbf{r}$ and $\eta$ are
component labels. Defining
\begin{equation}
\mathbf{\Phi}= \left(
\begin{array}{c} \Phi^{+}(\mathbf{r}) \\ \Phi^{-}(\mathbf{r}) \end{array}
\right),
\end{equation}
\begin{equation}
\mathbf{J}=   \left(
\begin{array}{r}
 \frac{e}{\hbar  } \sqrt{\frac{ 2}{\pi}}  E(\mathbf{r})\\
 J(\mathbf{r})
\end{array}
\right) ,
\end{equation}
\begin{equation} \mathsf{Q}= \frac{1}{\sqrt{2}} \left(
\begin{array}{cr}
1 & -1\\
1 & 1
\end{array}
\right) \, \delta(\mathbf{r}-\mathbf{r'}),
\end{equation}
and
\begin{equation}
\mathsf{C}_0= \left(
\begin{array}{cr}
\mathcal{C}^{++}_0(\mathbf{r},\mathbf{r'}) &
\mathcal{C}^{+-}_0(\mathbf{r},
\mathbf{r'}) \\
\mathcal{C}^{-+}_0(\mathbf{r},\mathbf{r'}) &
\mathcal{C}^{--}_0(\mathbf{r},
\mathbf{r'}) \\
\end{array}
\right) ,
\end{equation}
one can rewrite the generating functional (\ref{gen-fun-1}) as
%\begin{widetext}
\begin{eqnarray}
\displaystyle Z[J]&=& \frac{1}{\cal{N}_Z} \int {\mathcal D}
\mathbf{\Phi} \, e^{-\frac{1}{2} \left(\mathbf{\Phi}^T  {\mathsf
C}^{-1}_0 \mathbf{\Phi} -2 i \mathbf{J}^T  \mathsf{Q}
\mathbf{\Phi} \right)}
  \\ \nonumber
& \times&  \exp{\left\{ - \frac{i}{\hbar} \sum_{\eta=\pm} \eta
\int_{-\infty}^{+\infty} d t' \mathcal{H}_B[\Phi^{\eta}]
\right\}}\, ,
\end{eqnarray}
%\end{widetext}
where the superscript ${}^T$ indicates the transpose. Shifting the
fields
\begin{equation}
\mathbf{\Phi} \rightarrow \mathbf{\Phi}+{\mathbf{A}}_J\, ,
\hspace{1cm} {\mathbf{A}}_J=i{\mathbf{\mathsf{C}_0}}
{\mathsf{Q}}^T \mathbf{J} \, , \label{shift-in-field}
\end{equation}
the generating functional can be factorized into
\begin{equation}
Z[J]=Z_0[J]  Z_B[J] \; , \label{fun-gen-fact}
\end{equation}
where $Z_0$ and $Z_B$ are given below. In particular, $Z_0$ is the
generating functional in the absence of a backscatterer and reads
\begin{equation}
Z_0[J]= e^{-\frac{1}{2} \mathbf{J}^T \tilde{\mathsf C}_0
\mathbf{J}} \label{Z0}
\end{equation}
with
\begin{equation}
\tilde{\mathsf{C}}_0=\mathsf{Q}  \mathsf{C}_0
\mathsf{Q}^T=\left(\begin{array}{cc} 0 &
{\mathcal C}^{A}_0(\mathbf{r};\mathbf{r'})  \\
{\mathcal C}^{R}_0(\mathbf{r};\mathbf{r'}) &  {\mathcal
C}^{K}_0(\mathbf{r};\mathbf{r'})
\end{array} \right) \quad,
\end{equation}
where
\begin{eqnarray}
{\mathcal C}^{A}_0(\mathbf{r};\mathbf{r'}) &=& - \theta(t'-t)
\langle [ \Phi(\mathbf{r}), \Phi(\mathbf{r'}) ] \rangle_0 \; , \label{cadv} \\
{\mathcal C}^{R}_0(\mathbf{r};\mathbf{r'}) &=& \theta(t-t')
\langle [ \Phi(\mathbf{r}), \Phi(\mathbf{r'}) ] \rangle_0 \; ,
\label{cret}
\\ {\mathcal C}^{K}_0(\mathbf{r};\mathbf{r'}) &=& \langle \{
\Phi(\mathbf{r}),\Phi(\mathbf{r'})\} \rangle_0 \; \label{ckel} .
\end{eqnarray}
Exploiting the fact that ${\mathcal
C}^{A}_0(\mathbf{r'};\mathbf{r''})={\mathcal
C}^{R}_0(\mathbf{r''};\mathbf{r'})$, we can rewrite Eq.~(\ref{Z0})
as
\begin{widetext}
\begin{equation}
Z_0[J]= \exp{\left\{-\frac{1}{2} \int d\mathbf{r'} d\mathbf{r''}
J(\mathbf{r'}) {{\mathcal C}_0^K}(\mathbf{r'};\mathbf{r''})
J(\mathbf{r''}) -\frac{e}{\hbar} \sqrt{\frac{2}{\pi}} \int
d\mathbf{r'} d\mathbf{r''} J(\mathbf{r'}) {{\mathcal
C}_0^R}(\mathbf{r'};\mathbf{r''}) E(\mathbf{r''}) \right\} } \; .
\label{Z0bis}
\end{equation}
\end{widetext}
The second factor $Z_B$ in (\ref{fun-gen-fact}) is the generating
functional
\begin{eqnarray}\label{ZB}
 Z_B[J(\mathbf{r})] &&\\
 =&& \left \langle    \exp{\left(  -
\frac{i}{\hbar}  \sum_{\eta=\pm} \eta \int_{-\infty}^{+\infty}
\mathcal{H}_B[\Phi^{\eta}+A^{\eta}_J]  \, d t' \right)}
\right\rangle_0 \, , \nonumber
\end{eqnarray}
which  weights the backscattering term, and where the dependence
on the source field $J(\mathbf{x})$ is contained in the shift
${\mathbf{A}}_J$ defined in Eq.~(\ref{shift-in-field}). In
components, the latter reads explicitly
\begin{eqnarray}
A^{\eta}_J(\mathbf{r})&=& A_0(\mathbf{r})  \label{Aeta} \\
&+& \int \frac{d \mathbf{x}}{\sqrt{2}}
 \left[ i{\mathcal
C}^{K}_0(\mathbf{r};\mathbf{x})+\eta \, i{\mathcal
C}^{A}_0(\mathbf{r};\mathbf{x}) \right] J(\mathbf{x}) \,
,\nonumber
\end{eqnarray}
where
\begin{equation}
A_0(\mathbf{r})=\frac{e}{\sqrt{\pi} \hbar} \int d \mathbf{x} \,
i{\mathcal C}^{R}_0(\mathbf{r};\mathbf{x}) E(\mathbf{x})\, .
\label{A_0}
\end{equation}
Notice that $A_0$ is independent of $\eta$.

Using Eq.~(\ref{fun-gen-fact}), Eq.~(\ref{Phi-VM1}) can be
 rewritten as
\begin{equation}
\langle \Phi(\mathbf{x}) \rangle=\frac{-i}{\sqrt{2}} \left. \left(
\frac{\delta Z_0}{\delta J(\mathbf{x})} + \frac{\delta Z_B}{\delta
J(\mathbf{x})}\right) \right|_{J=0} \, . \label{Phi-VM2}
\end{equation}
It is now useful to define a quantity which has the dimension of a
current
\begin{equation}
j_{B}^\eta(\mathbf{r})=-\frac{e}{\hbar} \frac{\delta
\mathcal{H}_B}{\delta \Phi(\mathbf{r})}[\Phi^\eta+A^\eta_0] .
\end{equation}
Inserting Eqs.~(\ref{Z0bis}) and (\ref{ZB}) into
Eq.~(\ref{Phi-VM2}), and exploiting the property
\begin{displaymath}
\frac{\delta A^{\eta}_J (\mathbf{r}) }{ \delta J(\mathbf{x})} =
\frac{i \left( {\mathcal C}_0^K+\eta {\mathcal C}_0^A
\right)(\mathbf{r};\mathbf{x})}{\sqrt{2}}   = \frac{i \left(
{\mathcal C}_0^K+\eta {\mathcal C}_0^R
\right)(\mathbf{x};\mathbf{r})}{\sqrt{2}} ,
\end{displaymath}
we find
\begin{eqnarray}
\lefteqn{\langle \Phi(\mathbf{x}) \rangle =
 \frac{e}{\sqrt{\pi}\hbar}
  \int d\mathbf{r'}  \, i {{\mathcal
C}_0^R}(\mathbf{x};\mathbf{r'})  E(\mathbf{r'})} & &\label{Phi-MV3}   \\
& & + \frac{ 1}{2 e} \sum_{\eta=\pm} \int_{-\infty}^{+\infty} \!
\! \! dt' \left(\eta \, i {\mathcal C}^{K}_0 + \, i {\mathcal
C}^{A}_0\right)(\mathbf{r'_0};\mathbf{x}) \left\langle
j_{B}^\eta(\mathbf{r'_0}) \right\rangle_{\rightarrow} , \nonumber
\end{eqnarray}
where $\mathbf{r^\prime_0}=(x_0,t')$, ${\mathbf{x}}=(x,t)$, and
$\langle \ldots \rangle_{\rightarrow}$ denotes an average along
the Keldysh contour with respect to the shifted Hamiltonian
\begin{equation}
\mathcal{H}_{\rightarrow}=
\mathcal{H}_{0}[\Phi]+\mathcal{H}_{B}[\Phi+ A_0] \; . \label{S_0B}
\end{equation}
In the latter equation, $\mathcal{H}_{0}$ and $\mathcal{H}_{B}$
are given in Eqs.~(\ref{L0}) and (\ref{LB}), and the field in
$\mathcal{H}_{B}$ is shifted by $A_0$.

The current is now obtained from Eq.~(\ref{current}).
Differentiating Eq.~(\ref{Phi-MV3}) with respect to time and
defining
\begin{equation}
\sigma_0(\mathbf{x};\mathbf{y})=\frac{e^2}{h} 2 i \, \partial_t
{\mathcal C}^{R}_0(\mathbf{x};\mathbf{y}) \, , \label{sigma0def}
\end{equation}
one obtains
\begin{equation}
\langle j(\mathbf{x}) \rangle = I_0(\mathbf{x})-I_{\rm
BS}(\mathbf{x}) \label{Idef}
\end{equation}
with
\begin{equation}
I_0(\mathbf{x}) =  \int d \mathbf{r}'
\sigma_0(\mathbf{x};\mathbf{r}') E(\mathbf{r}') \quad , \, \,
\label{I0def}
\end{equation}
and
\begin{equation}
I_{\rm BS}(\mathbf{x}) = -\frac{\hbar \sqrt{\pi}}{e^2}
\int_{-\infty}^{+\infty} \! \! \! \! dt'
\sigma_0(\mathbf{x};{\mathbf{r}}^{\prime}_0) \left\langle
j_B^{+}({\mathbf{r}}^{\prime}_0) \right\rangle_{\rightarrow} .
\label{IBSdef}
\end{equation}
If we now specify  to the particular form (\ref{mu-deriv}), one
can easily show that Eq.~(\ref{A_0}) becomes
\begin{equation}
A_0(x,t)= A_0(t)= \frac{\omega_0 t}{2 \sqrt{\pi}} \; ,
\label{A_0_DC}
\end{equation}
where $\omega_0=e V/\hbar$ is the frequency related to the applied
voltage $V$ (see Eq.~(\ref{def-voltage})). In this case,
Eq.~(\ref{I0def}) gives $I_0(\mathbf{x})=e^2 V /h$ and
Eq.~(\ref{IBSdef}) becomes
\begin{widetext}
\begin{eqnarray}
I_{\rm BS}(\mathbf{x})  = -\frac{2\pi \lambda}{e}
\int_{-\infty}^{+\infty} dt' \,
\sigma_0(\mathbf{x};\mathbf{r^\prime_0}) \left\langle
\sin{[\sqrt{4 \pi} \Phi^+(\mathbf{r^\prime_0})+2k_F x_0+\omega_0
t' ]} \right\rangle_{\rightarrow} . \label{Phi-mv}
\end{eqnarray}
From an expansion of Eq.~(\ref{Phi-mv}) in terms of  $\lambda$, we
obtain for the average value of the backscattering current to
leading order in~$\lambda$
\begin{eqnarray}
I_{\rm BS}(\mathbf{x})  &=&  \frac{\pi \lambda^2}{2 e \hbar }
 \int_{-\infty}^{+\infty} dt' \,
\sigma_0(\mathbf{x};\mathbf{r^\prime_0}) \int_{-\infty}^{+\infty}
dt'' \sum_{m,m',n=\pm} m m' n \, e^{i (m+n) 2 k_F x_0 } e^{i
\omega_0 (m t''+n t')}
\nonumber \\
& &  \hspace{4cm} \times\left\langle e^{i m \sqrt{\pi}
(\Phi^+(\mathbf{r''_0})+\Phi^-(\mathbf{r''_0}))+ i m'\sqrt{\pi}
(\Phi^+(\mathbf{r''_0})-\Phi^-(\mathbf{r''_0}))+ i n  \sqrt{4\pi}
\Phi^+(\mathbf{r'_0})} \right\rangle_0   \nonumber
\\
&=&  \frac{\pi \lambda^2}{2 e \hbar }  \int_{-\infty}^{+\infty}
dt' \, \sigma_0(\mathbf{x};\mathbf{r^\prime_0})
\int_{-\infty}^{\infty} d t'' \, \sum_{m=\pm} m e^{i m \omega_0
(t''-t')} e^{4 \pi \langle \Phi(\mathbf{r''_0})
\Phi(\mathbf{r'_0}) -\Phi^2(\mathbf{r'_0}) \rangle_0 }   \nonumber \\
&=& \frac{e \lambda^{2}}{4 \hbar^2} \int_{-\infty}^{\infty} d t'
\, e^{i \omega_0 t'} \left( \sum_{s=\pm} s \, e^{4 \pi
{C}_0(x_0,st';x_0,0)} \right) ,
 \label{Phi-mv-2}
\end{eqnarray}
\end{widetext}
where
\begin{equation}
{C}_0(x_0,t'';x_0,t') \, = \,\langle \Phi(\mathbf{r''_0})
\Phi(\mathbf{r'_0}) -\Phi^2(\mathbf{r'_0}) \rangle_0 \; .
\end{equation}
One can thus see that $I_0$ and $I_{\rm BS}$ are actually
independent of $x$ and $t$, although the latter depends on the
impurity position $x_0$.

%%%%%%%%%%%%%%%%%%%%%%%%%%%%%%%%%%%%%%%%%%%%%%%%%%%%%%%%%%%%%%%%%%%%%%%%%%%%%%
%%%%%%%%%%%%%%%%%%%%%%%%%%%%%%%%%%%%%%%%%%%%%%%%%%%%%%%%%%%%%%%%%%%%%%%%%%%%%%
%%%%%%%%%%%%%%%%%%%%%%%%%%%%%%%%%%%%%%%%%%%%%%%%%%%%%%%%%%%%%%%%%%%%%%%%%%%%%%
%%%%%%     A P P E N D I X:  C O R R E L A T I O N   F U N C T I O N   %%%%%%%
%%%%%%%%%%%%%%%%%%%%%%%%%%%%%%%%%%%%%%%%%%%%%%%%%%%%%%%%%%%%%%%%%%%%%%%%%%%%%%
%%%%%%%%%%%%%%%%%%%%%%%%%%%%%%%%%%%%%%%%%%%%%%%%%%%%%%%%%%%%%%%%%%%%%%%%%%%%%%
%%%%%%%%%%%%%%%%%%%%%%%%%%%%%%%%%%%%%%%%%%%%%%%%%%%%%%%%%%%%%%%%%%%%%%%%%%%%%%

\section{Correlation Function}
\label{app_corr_fun} In this appendix, we derive an explicit
expression for  the  correlation function of the Bose field of the
ITLL model defined by the Hamiltonian (\ref{L0}).

We first observe that the field equation of motion obtained from
Eq.~(\ref{L0}) reads
\begin{equation}
\left(  \frac{1}{v_F^2} \frac{\partial^2}{\partial t^2}    -
\frac{\partial}{\partial x} \frac{1}{g^{2}(x)}
\frac{\partial}{\partial x} \right) \Phi(x,t) \, = \, 0 \, .
\label{eq-mot}
\end{equation}
The solution can easily be found by determining the eigenfunctions
of the inhomogeneous Laplacian, i.e., the $x$-dependent operator
on the l.h.s. of (\ref{eq-mot}). Denoting by $\Lambda$ the total
length of the system (the wire plus the leads), and imposing
periodic boundary conditions at $x=\pm \Lambda/2$ in the bulk of
the leads, the above eigenfunctions fall into two groups,
according to their parity ($S$=symmetric; $A$=antisymmetric), and
read respectively
\begin{widetext}
\begin{eqnarray}
 \psi_{S,k}(x)   =
\frac{1}{\sqrt{\Lambda}} \frac{1}{\sqrt{1+g^2-(1-g^2) \cos{(k g
L)}}}  \left\{
\begin{array}{ll}
\sum_{s=\pm} \, (g+s) \cos{[ k (x-\frac{L}{2}(1-s g))]}  & x
> \frac{L}{2}
\\
& \\
 2 g \cos{  (k g x) }
& |x| <  \frac{L}{2} \\ \\
\sum_{s=\pm} \, (g+s ) \cos{[  k (x+\frac{L}{2}(1-s g))]} & x <
-\frac{L}{2}
\end{array}
\right. \label{psi-sym}
\end{eqnarray}
\begin{eqnarray}
 \psi_{A,k}(x)   =
\frac{1}{\sqrt{\Lambda}} \frac{1}{\sqrt{1+g^2+(1-g^2) \cos{(k g
L)}}} \left\{
\begin{array}{ll}
\sum_{s=\pm} \, s(g+s) \sin{[ k (x-\frac{L}{2}(1-s g))]}  & x
> \frac{L}{2}
\\
& \\
 2 g \sin{  (k g x) }
& |x| <  \frac{L}{2}\label{psiA} \\ \\
\sum_{s=\pm} \, s(g+s) \sin{[  k (x+\frac{L}{2}(1-s g))]}   & x <
-\frac{L}{2}
\end{array}
\right. \label{psi-asym}
\end{eqnarray}
\end{widetext}
where  $k \in ]0; \infty[$ in the limit $\Lambda \rightarrow
\infty$. The related (doubly degenerate) eigenvalue reads $E_k=
\hbar \omega= \hbar v_F k$.

The (non-time ordered) correlation function for the unperturbed
system with Hamiltonian (\ref{L0}) can now easily be expressed in
terms of these eigenfunctions \cite{eco}, namely
\begin{eqnarray}
\lefteqn{\langle \Phi(x,t) \Phi(y,0)  \rangle_0 =} & & \label{corr-step1} \\
& & \sum_{b=S,A} \sum_{k
> 0} \frac{1}{2 k} \psi_{b,k}(x)
\psi_{b,k}(y) \left(e^{-i \omega t} + \frac{2 \cos{\omega t}
}{e^{\beta \hbar \omega}-1} \, \right) . \nonumber
\end{eqnarray}
As usual in one dimension, the expression (\ref{corr-step1}) needs
to be regularized, both in the infrared and ultraviolet regimes.
In order to avoid the infrared divergency, we introduce the
regularized correlation function
\begin{eqnarray}
\lefteqn{{C}_0(x,t;y,0)=} & & \nonumber \\
& & \left\langle \, \Phi(x,t) \Phi(y,0) -  \frac{\Phi^2(x,t) +
\Phi^2(y,0)}{2} \, \right\rangle_0 \label{Creg_def}
\end{eqnarray}
and, in order to avoid the ultraviolet divergency, we multiply the
r.h.s. of Eq.~(\ref{corr-step1}) by $e^{-\omega/\omega_c}$, where
the cut-off $\omega_c$ is related to the bandwidth.

From the knowledge of ${C}_0$ it is straightforward to compute the
advanced, retarded and Keldysh Green functions, respectively,
defined in Eqs.~(\ref{cadv})-(\ref{cret})-(\ref{ckel}), as well as
their Fourier transforms
\begin{equation} \label{cfourier}
\tilde{\mathcal{C}}_0^a(x,y,\omega)=\int_{-\infty}^{\infty} \,
e^{i \omega t} \, {\mathcal{C}}_0^a(x,t;y,0) \, dt \; ,
\end{equation}
for $a=A,R,K$.\\

It is useful to separate the contributions to ${C}_0$ coming from
the ground state (GS) and from thermal fluctuations (TF).
Accordingly, we split the average value appearing in
(\ref{Creg_def}) as $\langle \ldots \rangle \, = \, \langle \ldots
\rangle_{\rm GS} + \langle :\ldots: \rangle$, where the symbols
$\langle \ldots \rangle_{\rm GS}$ and $ : \, :$ respectively
indicate the average over the ground state of (\ref{L0}), and
Boson normal ordering. Thus, we have
\begin{equation}
{C}_{0} \, = \, {C}_{0;\rm GS}  \, + \, {C}_{0;\rm TF} \; .
\label{Creg}
\end{equation}
As one can see from Eq.~(\ref{corr-step1}), the temperature does
not enter in the imaginary part of ${C}_{0}$, and therefore the
conductivity $\sigma_0(x,y;\omega)$, related to ${C}_{0}$ through
Eqs.~(\ref{sig0omega}), (\ref{cfourier}), and (\ref{cret}), is
temperature independent. Below we explicitly give ${C}_0$ for $x$
and $y$ in the wire (i.e. $|x|,|y| \le L/2$), because this is the
case mostly needed in other sections.
\begin{widetext}
\begin{eqnarray}
\lefteqn{{C}_{0;\rm GS}(x,t;y,0)  =} \nonumber \\
& & -\frac{g}{4 \pi} \left\{ \sum_{m \in Z_{\rm even}}
\gamma^{|m|} \ln{\left(\frac{ (\alpha+i
\tau)^2+(\xi_r+m)^2}{\alpha^2+ m^2 }\right) } \, +
\right. \label{Creg_GS} \\
& & \hspace{1cm} + \left. \sum_{m \in Z_{\rm odd}} \gamma^{|m|}
\left\{ \ln{\left(\frac{ (\alpha+i
\tau)^2+(m-\xi_R)^2}{\alpha^2+(m-\xi_R)^2 }\right) } \, + \,
\frac{1}{2} \, \ln{ \left(\frac{[\alpha^2+(
\xi_R+m)^2]^2}{[\alpha^2+(2 \xi+m)^2] \,[\alpha^2+(2 \eta+m)^2]
}\right)}
\right\} \right\} \, , \nonumber \\
& & \nonumber \\
& & \nonumber \\
\lefteqn{{C}_{0;\rm TF}(x,t;y,0) = } \nonumber \\
& & -\frac{g}{4 \pi} \left[ \sum_{m \in Z_{\rm even}} \gamma^{|m|}
\ln \left( \frac{ |\Gamma{[1 + \Theta(\alpha+i m)]}|^4}{|\Gamma{[1
+ \Theta (\alpha+i (\tau+\xi_r+m))]}|^2 \,
  |\Gamma{[1 +
\Theta (\alpha+i (\tau-\xi_r-m))]}|^2} \right) \right. +
 \nonumber \\
& & \displaystyle \hspace{1.2cm}+ \!\!\! \sum_{m \in Z_{\rm odd}}
\gamma^{|m|} \ln \left( \frac{ |\Gamma{[1 + \Theta(\alpha+i
(m-\xi_R))]}|^4}{|\Gamma{[1 + \Theta (\alpha+i
(\tau+m-\xi_R))]}|^2 \,
  |\Gamma{[1 +
\Theta (\alpha+i (\tau-m+\xi_R))]}|^2}
 \right) + \nonumber  \\
& & \displaystyle \hspace{1.2cm}+ \left.   \sum_{m \in Z_{\rm
odd}} \gamma^{|m|} \ln   \frac{ |\Gamma{[1 + \Theta(\alpha+i(2
\xi+ m))]}|^2 \,  |\Gamma{[1 + \Theta(\alpha+i(2 \eta+
m))]}|^2}{|\Gamma{[1 + \Theta (\alpha+i (\xi_R+m))]}|^4 }
\right]\, , \label{Creg_TF}
\end{eqnarray}
\end{widetext}
where $\xi=x/L$, $\eta=y/L$, $\tau=t \omega_L$, $\Theta=k_B T/
\hbar \omega_L $, and $\alpha=\omega_L/\omega_c$ is the
(dimensionless) inverse cut-off. As one can see,
Eqs.~(\ref{Creg_GS}) and (\ref{Creg_TF}) depend on time and
temperature only through $t \omega_L$ and $k_B T/\hbar \omega_L$.

The part of the correlation function coming from thermal
fluctuations does actually not need an ultraviolet cutoff, because
the role of the cutoff is equivalently played by the finite
temperature. Therefore, we can send $\alpha \rightarrow 0$ in
Eq.~(\ref{Creg_TF}) and obtain the more familiar form
\begin{widetext}
\begin{eqnarray}
\lefteqn{{C}_{0;\rm TF}(x,t;y,0) = } \nonumber \\
& & -\frac{g}{4 \pi} \left[ \sum_{m \in Z_{\rm even}} \gamma^{|m|}
\sum_{r=\pm} \,\ln \left( \frac{\sinh{[\pi\Theta (\tau+r(\xi_r+m))
]}  }{\pi \Theta (\tau+r(\xi_r+m)) }  \frac{\pi \Theta m}{
\sinh{[\pi \Theta m]} } \right) \right. +
 \nonumber \\
& & \displaystyle \hspace{1.0cm}+ \!\!\! \sum_{m \in Z_{\rm odd}}
\gamma^{|m|}  \sum_{r=\pm} \, \ln \left( \frac{\sinh{[\pi \Theta
(\tau+r(m-\xi_R)) ]}}{\pi \Theta(\tau+r(m-\xi_R))}  \,
 \frac{\pi \Theta (m-\xi_R)}{\sinh{[\pi \Theta  (m-\xi_R)]}}
 \right) + \nonumber  \\
& & \displaystyle \hspace{1.0cm}+ \left. \sum_{m \in Z_{\rm odd}}
\gamma^{|m|} \ln \left(\frac{\sinh^2{[\pi \Theta (\xi_R+m)
]}}{[\pi \Theta(\xi_R+m)]^2}  \frac{\pi \Theta (2 \xi + m) } {
\sinh{[\pi \Theta(2 \xi + m)]}}  \frac{\pi \Theta (2 \eta + m) } {
\sinh{[\pi \Theta(2 \eta+ m)]}}\right)
 \right] \; , \label{Creg_TF2}
\end{eqnarray}
\end{widetext}
where we used the relation
\begin{equation}
|\Gamma[1+i X]|^2 = \frac{\pi X}{\sinh \pi X} \; ,
\label{Gamma1+iX}
\end{equation}
which holds if $X$ is real.\\
%%%%%%%%%%%%%%%%%%%%%%%%%%%%%%%%%%%%%%%%%%%%%%%%%%%%%%%%%%%%%%%%%%%%%%%%%%%%%%
%%%%%%%%%%%%%%%%%%%%%%%%%%%%%%%%%%%%%%%%%%%%%%%%%%%%%%%%%%%%%%%%%%%%%%%%%%%%%%
%%%%%%%%%%%%%%%%%%%%%%%%%%%%%%%%%%%%%%%%%%%%%%%%%%%%%%%%%%%%%%%%%%%%%%%%%%%%%%
%%%%%%    A P P E N D I X:   A S Y M P T O T I C   E X P A N S I O N    %%%%%%
%%%%%%%%%%%%%%%%%%%%%%%%%%%%%%%%%%%%%%%%%%%%%%%%%%%%%%%%%%%%%%%%%%%%%%%%%%%%%%
%%%%%%%%%%%%%%%%%%%%%%%%%%%%%%%%%%%%%%%%%%%%%%%%%%%%%%%%%%%%%%%%%%%%%%%%%%%%%%
%%%%%%%%%%%%%%%%%%%%%%%%%%%%%%%%%%%%%%%%%%%%%%%%%%%%%%%%%%%%%%%%%%%%%%%%%%%%%%
\section{Asymptotic expansion}
\label{app_coeff} 
\begin{figure}
\vspace{0.3cm}
\begin{center}
\epsfig{file=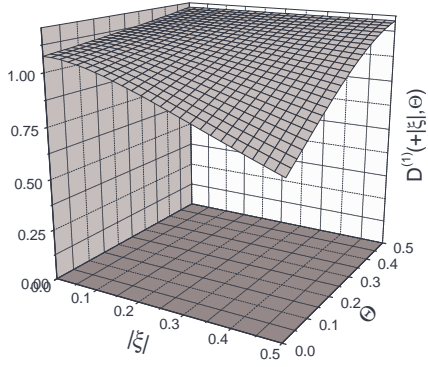,scale=0.3} \caption{\label{fig_coeff1}
The coefficient ${D}^{(1)}(+|\xi|,\Theta)$ as a function of the
relative impurity position $\xi$  and the dimensionless
temperature $\Theta=k_B T / \hbar \omega_L$ for $g=0.25$.}
\end{center}
\end{figure}

In the present appendix, we provide the explicit
expressions for the coefficients ${D}^{(1)}(\pm|\xi|;\Theta)$ and
${D}^{(2)}(\Theta)$ appearing in the asymptotic expansions
(\ref{asy-exp-xneq0}) and (\ref{asy-exp-xeq0}). The former can be
written as a product of a GS and a TF contribution
\begin{equation}
{D}^{(1)}(\pm|\xi|;\Theta) \, = \, {D}^{(1)}_{\rm GS} (\pm|\xi|)
{D}^{(1)}_{\rm TF} (\pm|\xi|;\Theta) \; , \label{D^1}
\end{equation}
where
\begin{widetext}
\begin{eqnarray}
{D}^{(1)}_{\rm GS}(\pm|\xi|) \,& =& \, \left(\frac{4}{3\mp 2
|\xi|}\right)^{2 g \gamma^2}  \left( \frac{(9-4|\xi|^2)^2}{64
(1\mp |\xi|) }\right)^{g
\gamma^3}   \nonumber \\
 & &   \nonumber \\
& \times & \prod_{p =2}^{\infty} \, \left\{
 \left(\frac{ (2 p)^2 }{ (2 p)^2-(1\mp 2
|\xi|)^2}\right)^{2 g \gamma^{2p}}  \prod_{r=\pm} \left(\frac{
(2p+1+2 r |\xi|)^2 }{(2p+1+2 r |\xi|)^2-(1 \mp
2|\xi|)^2}\right)^{g \gamma^{2 p+1}} \right\} \label{coeff-D_GS}
\end{eqnarray}
and
\begin{eqnarray}
{D}^{(1)}_{\rm TF}(\pm|\xi|;\Theta) &=& \left(
 \frac{ \pi \Theta  (1\mp 2 |\xi|)]}{\sinh[ \pi \Theta( 1\mp 2 |\xi|)]}
 \right)^{2g} \,  \prod_{p =1}^{\infty}
 \prod_{s=\pm} \left\{ \left(\frac{\pi \Theta  (1 \mp
2|\xi| + 2 p s) }{ \sinh[\pi \Theta  (1 \mp 2|\xi| + 2 p s) ]}
\frac{\sinh[\pi \Theta 2p ]}{ \pi \Theta  2p }
 \right)^{g\gamma^{2p}}   \right. \label{coeff-D_TF} \\
& \times&   \left.   \prod_{r=\pm} \left(\frac{\pi \Theta  (1 \mp
2|\xi| + s (2p-1+2r |\xi|)) }{ \sinh[\pi \Theta  (1 \mp 2|\xi| + s
(2p-1+2r |\xi|)) ]}  \frac{\sinh[\pi \Theta (2p-1+2 r|\xi|)]}{ \pi
\Theta (2p-1+2 r|\xi|)}
 \right)^{g\gamma^{2p-1}}
 \right\}  \, . \nonumber
\end{eqnarray}
\end{widetext}
In Fig.~\ref{fig_coeff1}, we have plotted
${D}^{(1)}(+|\xi|,\Theta)$ for $g=0.25$ as a function of the
relative impurity position $|\xi_0| \in [0 , 1/2]$ and the
temperature $\Theta=k_B T/ \hbar \omega_L$, in the range $\Theta
\in [0 , 1/2]$. Note that the expression (\ref{asy-exp-xneq0}) is
by definition valid for $\Theta \ll 1$ only. As one can see, the
coefficient has a maximum at $\Theta=0$, which is of order $1$.
Similar values are also obtained for ${D}^{(1)}(-|\xi|,\Theta)$,
and for other values of the interaction strength $g$.
%%%%%%%%%%%%%%%%%%%%%%%%%%%%%%%%%%%%%%%
%%%%%%%%%%%%%%%%%%%%%%%%%%%%%%%%%%%%%%%
%%%%%%%%%%%%%%%%%%%%%%%%%%%%%%%%%%%%%%%
%%%%%      FIGURE  Coeff 1       %%%%%%
%%%%%%%%%%%%%%%%%%%%%%%%%%%%%%%%%%%%%%%
%%%%%%%%%%%%%%%%%%%%%%%%%%%%%%%%%%%%%%%

\begin{figure}
\vspace{0.3cm}
\begin{center}
\epsfig{file=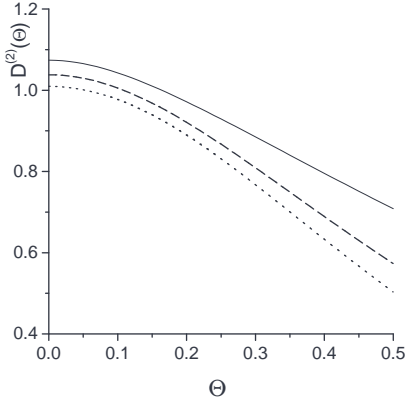,scale=0.3} \caption{\label{fig_coeff2} The
coefficient ${D}^{(2)}(\Theta)$ as a function of dimensionless
temperature $\Theta=k_B T / \hbar \omega_L$ for $g=0.25$ (solid
line), $g=0.5$ (dashed line), and $g=0.75$ (dotted line).}
\end{center}
\end{figure}

The coefficient ${D}^{(2)}(\Theta)$ in Eq.~(\ref{asy-exp-xeq0})
reads
\begin{equation}
{D}^{(2)}( \Theta) \, = \, {D}^{(2)}_{\rm GS}   {D}^{(2)}_{\rm TF}
(\Theta) \; , \label{D^2}
\end{equation}
where
\begin{equation}
{D}^{(2)}_{\rm GS} \, = \,   \prod_{m =2}^{\infty} \,
 \left(\frac{ m^2  }{ m^2-1}\right)^{2 g \gamma^{|m|}}
\end{equation}
and
\begin{equation}
{D}^{(2)}_{\rm TF}(\Theta) = \prod_{m =-\infty}^{\infty}  \left(
\frac{\pi \Theta (1+m) }{\sinh[\pi \Theta (1+m) ]} \frac{\sinh[\pi
\Theta m ]}{\pi \Theta m }
 \right)^{2 g \gamma^{|m|}} .
\end{equation}
This coefficient is shown in Fig.~\ref{fig_coeff2} for three
different values of the interaction strength $g$.
%%%%%%%%%%%%%%%%%%%%%%%%%%%%%%%%%%%%%%%
%%%%%      FIGURE  Coeff 2       %%%%%%
%%%%%%%%%%%%%%%%%%%%%%%%%%%%%%%%%%%%%%%
%%%%%%%%%%%%%%%%%%%%%%%%%%%%%%%%%%%%%%%
\\
%%%%%%%%%%%%%%%%%%%%%%%%%%%%%%%%%%%%%%%%%%%%%%%%%%%%%%%%%%%%%%%%%%%%%%%%%%%%%%
%%%%%%%%%%%%%%%%%%%%%%%%%%%%%%%%%%%%%%%%%%%%%%%%%%%%%%%%%%%%%%%%%%%%%%%%%%%%%%
%%%%%%%%%%%%%%%%%%%%%%%%%%%%%%%%%%%%%%%%%%%%%%%%%%%%%%%%%%%%%%%%%%%%%%%%%%%%%%
%%%%%%    A P P E N D I X:    F I N I T E  F R E Q U E N C Y  N O I S E %%%%%%
%%%%%%%%%%%%%%%%%%%%%%%%%%%%%%%%%%%%%%%%%%%%%%%%%%%%%%%%%%%%%%%%%%%%%%%%%%%%%%
%%%%%%%%%%%%%%%%%%%%%%%%%%%%%%%%%%%%%%%%%%%%%%%%%%%%%%%%%%%%%%%%%%%%%%%%%%%%%%
%%%%%%%%%%%%%%%%%%%%%%%%%%%%%%%%%%%%%%%%%%%%%%%%%%%%%%%%%%%%%%%%%%%%%%%%%%%%%%

\section{Local conductivity and noise} \label{app_noise}
\noindent The local conductivity  is defined as
\begin{equation}
\sigma(x,y,\omega)= \int_{-\infty}^{+\infty} dt \ e^{i \omega t}
\sigma(x,t;y,0)  ,
\end{equation}
where
\begin{equation}
\sigma(\mathbf{x};\mathbf{y})= \left. \frac{\delta \langle
j(\mathbf{x}) \rangle}{\delta E(\mathbf{y})} \right|_{E = 0} .
\end{equation}
Here, $\langle j(\mathbf{x}) \rangle$ is given by
Eq.~(\ref{Idef}), and $E$ is the external source appearing in the
Hamiltonian (\ref{LE}). Performing the functional derivative, and
making use of the identity
\begin{widetext}
\begin{eqnarray}
\left\langle \frac{\delta^2 \mathcal{H}_B}{\delta
\Phi^2(\mathbf{r'_0})}[\Phi^{\eta}+A^{\eta}_0]
\right\rangle_{\rightarrow} = \frac{i}{\hbar}
\int_{-\infty}^{\infty} dt'' \sum_{\eta_2=\pm} \eta_2 \left\langle
\frac{\delta \mathcal{H}_B}{\delta
\Phi(\mathbf{r'_0})}[\Phi^{\eta}+A^{\eta}_0]  \frac{\delta
\mathcal{H}_B}{\delta
\Phi(\mathbf{r''_0})}[\Phi^{\eta_2}+A^{\eta_2}_0]
\right\rangle_{\rightarrow} \, ,\label{crucial-identity}
\end{eqnarray}
one  easily obtains  for the non-local conductivity
\begin{eqnarray}
  \sigma(x,y,\omega) = \sigma_0(x,y,\omega)-
  \frac{1}{2 h \omega} \left( \frac{h}{e^2}
\right)^2 \sigma_0(x,x_0,\omega) \sigma_0(x_0,y,\omega)  \left.
f_C(x_0,\omega) \right|_{E=0} \; , \label{sigmadef}
\end{eqnarray}
where $\sigma_0(x,y,\omega)$ is given by Eqs.~(\ref{sig0omega})
and (\ref{cfourier}), and $f_C$ by Eq.~(\ref{s_C}).
\end{widetext}
The  noise is computed starting from the expression
\begin{eqnarray}
\langle \{\Phi(\mathbf{x}),\Phi(\mathbf{y})\} \rangle  &=&
 \sum_{\eta,\eta'} \langle
\frac{\Phi^\eta(\mathbf{x}) \Phi^{\eta'}(\mathbf{y})}{2}
\rangle \nonumber \\
&=& - \left. \frac{\delta^2 Z[J]}{\delta J(\mathbf{x}) \delta
J(\mathbf{y})} \right|_{J=0} \; , \label{Phi-fluct}
\end{eqnarray}
which, using Eq.~(\ref{fun-gen-fact}),  can be rewritten as
\begin{eqnarray}
&& \langle \{ \Phi(\mathbf{x}), \Phi(\mathbf{y}) \} \rangle -2
\langle \Phi(\mathbf{x}) \rangle \langle \Phi(\mathbf{y}) \rangle
= \label{fluctu} \\
&& \sum_{a=0,B} \left. \left( \frac{\delta Z_a}{\delta
J(\mathbf{x})}   \frac{\delta Z_a}{\delta J(\mathbf{y})}
 - \frac{\delta^2 Z_a}{\delta
J(\mathbf{x})\delta J(\mathbf{y})} \right) \right|_{J=0} \; .
\nonumber
\end{eqnarray}
In particular, from Eq.~(\ref{Z0bis}) we obtain
\begin{equation}
\left. \frac{\delta Z_0}{\delta J(\mathbf{x})} \frac{\delta
Z_0}{\delta J(\mathbf{y})} - \frac{\delta^2 Z_0}{\delta
J(\mathbf{x}) \delta J(\mathbf{y})} \right|_{J=0}=
{\mathcal{C}}_0^K(\mathbf{x};\mathbf{y}) \; ,\label{Z0intero}
\end{equation}
while the backscattering part (\ref{ZB}) gives
\begin{widetext}
\begin{eqnarray}
\left. \frac{\delta^2 Z_B}{\delta J(\mathbf{x}) \delta
J(\mathbf{y})} \right|_{J=0}&=& \frac{i}{2\hbar}
\int_{-\infty}^{\infty} dt' \sum_{\eta_1=\pm} \left( \eta_1
{\mathcal{C}}_0^K+{\mathcal{C}}_0^R \right)\! \!
(\mathbf{x};\mathbf{r'_0}) \left({\mathcal{C}}_0^K+\eta_1
\mathcal{C}_0^R \right) \! \!(\mathbf{y};\mathbf{r'_0})
\left\langle \frac{\delta^2 \mathcal{H}_B}{\delta
\Phi^2(\mathbf{r'_0})}[\Phi^{\eta_1}+A_0]
\right\rangle_{\rightarrow}
\nonumber \\
& &+\frac{1}{2\hbar^2} \int_{-\infty}^{\infty} dt'
\int_{-\infty}^{\infty} dt'' \sum_{\eta_1=\pm} \sum_{\eta_2=\pm}
\left( \eta_1 {\mathcal{C}}_0^K+{\mathcal{C}}_0^R \right)\! \!
(\mathbf{x};\mathbf{r'_0}) \left(
\eta_2{\mathcal{C}}_0^K+{\mathcal{C}}_0^R \right)\! \!
(\mathbf{y};\mathbf{r''_0})   \nonumber \\
& & \hspace{4cm} \times \left\langle \frac{\delta
\mathcal{H}_B}{\delta \Phi(\mathbf{r'_0})}[\Phi^{\eta_1}+A_0]
\frac{\delta \mathcal{H}_B}{\delta
\Phi(\mathbf{r''_0})}[\Phi^{\eta_2}+A_0]
\right\rangle_{\rightarrow} ,
\end{eqnarray}
where $\mathbf{r''_0}=(x_0,t'')$. Observing that $\langle
\frac{\delta^2 \mathcal{H}_B}{\delta \Phi^2}[\Phi^{\eta}+A_0]
\rangle_{\rightarrow}$ is independent of $\eta$, and exploiting
the identity (\ref{crucial-identity}), one obtains
\begin{eqnarray}
 & & \left. \frac{\delta^2 Z_B}{\delta J(\mathbf{x})
\delta J(\mathbf{y})} \right|_{J=0} = -\frac{1}{2 e^2} \int \! \!
\! \! \int dt'   dt'' \! \! \! \sum_{\eta_1,\eta_2=\pm} \left(
{\mathcal{C}}_0^K \! \!(\mathbf{x};\mathbf{r'_0})
 {\mathcal{C}}_0^R
 (\mathbf{y};\mathbf{r'_0})
 +{\mathcal{C}}_0^R(\mathbf{x};\mathbf{r'_0})
\mathcal{C}_0^K (\mathbf{y};\mathbf{r'_0}) \right) \,
 \eta_2 \! \left\langle
j_{B}^{\eta_1}(\mathbf{r'_0}) j_{B}^{\eta_2}(\mathbf{r''_0})
\right\rangle_{\rightarrow}
   \nonumber \\
& & +\frac{1}{2 e^2} \int \! \! \! \! \int dt'   dt''
\sum_{\eta_1,\eta_2=\pm} \left\{
{\mathcal{C}}_0^K(\mathbf{x};\mathbf{r'_0}) \mathcal{C}_0^R
(\mathbf{y};\mathbf{r''_0}) \,   \eta_1 \! \left\langle
j_{B}^{\eta_1}(\mathbf{r'_0}) j_{B}^{\eta_2}(\mathbf{r''_0})
 \right\rangle_{\rightarrow} \, + \,   {\mathcal{C}}_0^R(\mathbf{x};
  \mathbf{r'_0})
\mathcal{C}_0^K (\mathbf{y};\mathbf{r''_0})
 \,  \eta_2 \! \left\langle
j_{B}^{\eta_1}(\mathbf{r'_0}) j_{B}^{\eta_2}(\mathbf{r''_0})
\right\rangle_{\rightarrow} \right.  \nonumber \\
& & \,  \hspace{6cm}  \left. + \,
{\mathcal{C}}_0^R(\mathbf{x};\mathbf{r'_0}) \mathcal{C}_0^R
(\mathbf{y};\mathbf{r''_0})    \left\langle
j_{B}^{\eta_1}(\mathbf{r'_0}) j_{B}^{\eta_2}(\mathbf{r''_0})
\right\rangle_{\rightarrow}  \right\} . \label{ZBpezzo1}
\end{eqnarray}
\end{widetext}
Furthermore, we have
\begin{eqnarray}
 \left. \frac{\delta Z_B}{\delta J(\mathbf{x})}
\frac{\delta Z_B}{\delta J(\mathbf{y})} \right|_{J=0} =
\label{ZBpezzo2}
\hspace{4.5cm} \\
  \frac{2}{e^2} \int \! \! \! \! \int dt'   dt''
{\mathcal{C}}_0^R(\mathbf{x};\mathbf{r'_0}) \mathcal{C}_0^R
(\mathbf{y};\mathbf{r''_0}) \left\langle
j_{B}^{\eta}(\mathbf{r'_0})
 \right\rangle_{\rightarrow}  \left\langle j_{B}^{\eta}(\mathbf{r''_0})
 \right\rangle_{\rightarrow}
 . \nonumber
\end{eqnarray}
From the general definition (\ref{noise}) of the noise and the
expression (\ref{current}) for the current we obtain
\begin{eqnarray}
S(x,y,\omega)&=&\frac{e^2}{\pi} \! \! \int_{-\infty}^{\infty} \!
\! \! e^{i \omega t}  \left( \langle \{\dot{\Phi}(x,t)
\dot{\Phi}(y,0) \} \rangle - \right.  \label{S-prelim} \\ & &
\hspace{2cm}- \left. 2
 \langle \dot{\Phi}(x,t) \rangle \langle \dot{\Phi}(y,0)  \rangle \right) .
 \nonumber
\end{eqnarray}
This formula for the noise  can now be computed by inserting
Eq.~(\ref{fluctu}) and using Eqs.~(\ref{Z0intero}),
(\ref{ZBpezzo1}), and (\ref{ZBpezzo2}). With the help of the
relations
%\begin{widetext}
\begin{eqnarray*}
&& \sum_{\eta_1,\eta_2=\pm} \hspace{-0.3cm} \eta_1 \left\langle
j_{B}^{\eta_1}(\mathbf{r'}) j_{B}^{\eta_2}(\mathbf{r''})
\right\rangle_{\rightarrow} = \\
&& - 2 \theta(t''-t') \left\langle [ j_{B}(\mathbf{r'}) ,
j_{B}(\mathbf{r''}) ] \right\rangle_{\rightarrow} ,
\end{eqnarray*}
\[
\sum_{\eta_1,\eta_2=\pm} \hspace{-0.3cm} \eta_2 \left\langle
j_{B}^{\eta_1}(\mathbf{r'}) j_{B}^{\eta_2}(\mathbf{r''})
\right\rangle_{\rightarrow}  = 2 \theta(t'\!-\!t''\!) \left\langle
[ j_{B}(\mathbf{r'}) , j_{B}(\mathbf{r''}) ]
\right\rangle_{\rightarrow} ,
\]
\[
\sum_{\eta_1,\eta_2=\pm} \hspace{-0.3cm}\left\langle
j_{B}^{\eta_1}(\mathbf{r'}) j_{B}^{\eta_2}(\mathbf{r''})
\right\rangle_{\rightarrow} = 2 \left\langle \{ j_{B}(\mathbf{r'})
, j_{B}(\mathbf{r''}) \} \right\rangle_{\rightarrow} ,
\]
and
\[
\sum_{\eta_1,\eta_2=\pm} \hspace{-0.3cm} \eta_1 \eta_2 \left
\langle j_{B}^{\eta_1}(\mathbf{r'}) j_{B}^{\eta_2}(\mathbf{r''})
\right\rangle_{\rightarrow}     =   0 ,
\]
%\end{widetext}
we derive the final result
\begin{equation} \label{s_result}
S(x,y,\omega)=S_0(x,y,\omega)+S_A(x,y,\omega)+S_C(x,y,\omega) \; ,
\label{SABC}
\end{equation}
where the three contributions to the FF noise read
\begin{equation}
S_0(x,y,\omega) = \frac{e^2 \omega^2}{\pi}
\tilde{\mathcal{C}}_0^K(x,y,\omega) , \label{S0app}
\end{equation}
\begin{widetext}
\begin{equation}
S_A(x,y,\omega) = -\frac{  \omega^2}{\pi}
\tilde{\mathcal{C}}_0^R(x,x_0,\omega)
\tilde{\mathcal{C}}_0^R(y,x_0,-\omega) f_A(x_0,\omega) ,
\label{SAapp}
\end{equation}
and
\begin{equation}
S_C(x,y,\omega)=  -\frac{ \omega^2}{\pi} \left(
\tilde{\mathcal{C}}_0^K(x,x_0,\omega)
\tilde{\mathcal{C}}_0^R(y,x_0,-\omega) f_C(x_0,-\omega)+
\tilde{\mathcal{C}}_0^R(x,x_0,\omega)
\tilde{\mathcal{C}}_0^K(y,x_0,-\omega) f_C(x_0,\omega) \right)
\quad. \label{SCapp}
\end{equation}
\end{widetext}
%
%\bibliography{data}

\begin{thebibliography}{10}
\bibitem{NOTA} This can, for instance, be seen from Eq.~(\ref{A_0}).
If the correlation function of the homogenous TLL were used
instead of $i{\mathcal C}^{R}_0(\mathbf{r};\mathbf{x})$, the
result of Eq.~(\ref{A_0_DC}) would have a prefactor $g$ in front
of it. Essentially, the homogeneous TLL performes the
time-integral in Eq.~(\ref{A_0}) over times that are always
smaller than the traversal time $\omega_L^{-1}$ of the wire,
whereas the ITLL does not include such a restriction.
\bibitem{NOTA2}
Although the period of the function $F(\omega)$ in
Eq.~(\ref{fofomega}) is $2\pi\omega_L=2 \Delta \omega_1$, by
virtue of the symmetry of $F(\omega)$ around $\omega=\pi \omega_L$
(see Fig.~\ref{Fano}), an averaging over half the period is
sufficient to obtain the result in Eq.~(\ref{s_ex_average}).
\end{thebibliography}
%\bibliographystyle{prsty}

\end{document}